\documentclass[draft]{agujournal2019}

\usepackage{amsmath}
\usepackage{amssymb}

\usepackage{algorithm}
\usepackage{algpseudocode}

\usepackage{graphicx}
\usepackage{tabularx}
\usepackage{booktabs}

\usepackage{xcolor}
\usepackage{soul}
\soulregister\ref7
\soulregister\cite7
\soulregister\citeA7
\soulregister\eqref7
\usepackage{url}

\usepackage[symbol]{footmisc}

\journalname{Journal of Advances in Modeling Earth Systems (JAMES)}
\begin{document}

\title{Gradient-free online learning of subgrid-scale dynamics with neural emulators}

\authors{H. Frezat\affil{1,2,3,4}\footnote{https://hrkz.github.com}, R. Fablet\affil{3,4}, G. Balarac\affil{1,5}, and J. Le Sommer\affil{2}}

\affiliation{1}{Univ. Grenoble-Alpes, CNRS, Grenoble INP, LEGI, Grenoble, France}
\affiliation{2}{Univ. Grenoble-Alpes, CNRS, INRAE, IRD, Grenoble INP, IGE, Grenoble, France}
\affiliation{3}{IMT Atlantique, CNRS, Lab-STICC, Brest, France}
\affiliation{4}{Inria, Odyssey team, Brest, France}
\affiliation{5}{Institut Universitaire de France (IUF), Paris, France}

\correspondingauthor{Hugo Frezat}{hugo.frezat@gmail.com}

\date{October 2024}

\begin{keypoints}
\item Parametrizations learned with online approaches are known to be more stable but require differentiable solvers.
\item We train a differentiable emulator of the coarse-resolution solver that enables online approaches for the parametrization.
\item This opens new avenues to learn stable parametrizations for non-differentiable legacy codes.
\end{keypoints}

\begin{abstract}
In this paper, we propose a generic algorithm to train machine learning-based subgrid parametrizations online, i.e., with \textit{a posteriori} loss functions, but for non-differentiable numerical solvers. The proposed approach leverages a neural emulator to approximate the reduced state-space solver, which is then used to allow gradient propagation through temporal integration steps. We apply this methodology on a chaotic two-timescales Lorenz-96 system and a single layer quasi-geostrophic system with zonal dynamics, known to be highly unstable with offline strategies. Using our algorithm, we are able to train a parametrization that recovers most of the benefits of online strategies without having to compute the gradient of the original solver. We found that training the neural emulator and parametrization components separately with different loss quantities is necessary in order to minimize the propagation of approximation biases. Experiments on emulator architectures with different complexities also indicates that emulator performance is key in order to learn an accurate parametrization. This work is a step towards learning parametrization with online strategies for climate models.
\end{abstract}

\section*{Plain Language Summary}
We introduce a new way to train machine-learning models that are designed to fill in the ``missing physics" in climate simulations. These models, called parametrizations, help large-scale solvers represent small-scale processes that are too expensive to compute directly. Ideally, these parametrizations should be trained ``online", meaning they learn while interacting with the numerical solver. However, climate solvers use numerical methods that do not allow gradients to pass through them, which makes standard online training techniques impossible. Our approach overcomes this limitation by building a separate ``emulator" that imitates the behavior of the original simulation. This emulator is smooth and differentiable, which allows us to compute gradients and train the parametrization as if it was interacting with the real solver. We test this approach on two simplified systems that share features with atmospheric dynamics, including chaotic behavior and fast–slow interactions. We show that our method produces parametrizations that closely match the performance of true online training, without requiring gradients from the original solver. We also find that training the emulator and the parametrization with different objective ``measures" reduces bias and improves accuracy. This approach brings us closer to using online learning in full-scale climate systems.

\section{Introduction}
Numerical models used for simulating the evolution of fluid flows generally require subgrid scales (SGS) and physical parameterizations.
These serve the purpose of representing the impact of scales that remain unresolved by the discretized solver as for instance subgrid turbulent processes \cite{rogallo1984numerical,meneveau2000scale}. 
They may also account for other physical phenomena that are not explicitly accounted for by the partial differential equations.
SGS and physical parameterizations are generally considered to be key components of numerical models in Earth System science \cite{fox2019challenges,schneider2017earth}. 
In these numerical models, they are often rate-controlling components which largely govern the long term behavior of model solutions \cite{maher2018impact,hewitt2020resolving}. 
This is why significant efforts are concentrated on the development of SGS and physical parameterizations in these communities. %
Machine learning (ML) techniques have now been widely applied to SGS and physical parametrizations for numerical simulations, as already discussed in reviews \cite{brunton2020machine, vinuesa2022enhancing}. These ML-based components have been demonstrated in different applications, including atmospheric circulation \cite{rasp2018deep,yuval2020stable}, oceanic circulation \cite{bolton2019applications,guillaumin2021stochastic}, and sea-ice dynamics \cite{finn2023deep} but also with benchmarks of toy models \cite{ross2023benchmarking}. The ambition of this emerging field is to design hybrid numerical models that improves upon existing ones by combining physics-based partial differential equation (PDE) solvers and ML-based trainable components. In this framework, well-constrained aspects are incorporated into the resolved equations, while components that are less rigorously constrained or follow a statistical behavior are represented by ML-based algorithms. When possible, physical knowledge can also be incorporated in the ML-based component \cite{pawar2023frame,frezat2021physical,beucler2021enforcing}.
Hybrid numerical models are currently designed through an optimization procedure that attempts to improve its ability to replicate the results of a higher fidelity model. In practice, higher fidelity means that it can reproduce additional physical processes (e.g., microphysics) \cite{seifert2020potential}, contains information from reanalyses \cite{mcgibbon2019single,watt2021correcting} or from higher resolutions \cite{frezat2022posteriori}. The hybrid numerical models are typically run at coarser resolution and the parameters of their ML components are optimized in order to fit a given training objective. This objective is usually formulated from pre-existing high-resolution, process resolving numerical models (i.e. direct numerical simulation (DNS), or high-resolution version of the same code) but also from observations or lab experiments. By essence, this optimization is very much related to the tuning of model parameters, or to the estimation of state-dependent correction of models.

Several recent results suggest that ML-based SGS parametrization give better results when the ML components are trained with online strategies \cite{sirignano2020dpm,macart2021embedded,duraisamy2021perspectives,sirignano2023dynamic,pahlavan2024explainable}. Similarly, \cite{kochkov2024neural} developed a neural general circulation model with online-trained parametrizations based on ERA5 data. The training of ML components for hybrid numerical models is said to use an online strategy if the evaluation of the training objective involves integrating the hybrid numerical model for several time steps. Note that the same idea has also been referred to as \textit{a posteriori} \cite{frezat2022posteriori}, solver-in-the-loop \cite{list2022learned}, differentiable \cite{shankar2023differentiable,fan2024differentiable} or end-to-end in the ML community. In practice, the strategy has been shown to lead to more stable simulations, with better performance in \textit{a posteriori} tests. However, the online strategy typically requires to compute the derivative of the hybrid numerical model with respect to the parameters of the ML components.
These results suggest that, as in other fields of physical science, the emerging paradigm of differentiable programming (DP) shows great potential for the design of hybrid numerical models applied to the simulations of turbulent systems.
DP is often combined with automatic differentiation (AD) which allows to automatically obtain the derivatives of a chain of operations in order to be used in an optimization procedure based on a gradient descent variant. AD frameworks are typically implemented by specific libraries or programming languages, and have seen many new development applications in physical sciences \cite{um2020solver, holl2020learning, heiden2021neuralsim, negiar2023learning}. There has been a particularly active community around the problem of turbulence modeling \cite{dresdner2023learning, kochkov2021machine, takahashi2021differentiable} and climate-geoscientific modeling \cite{gelbrecht2023differentiable, wagner2023catke, ramadhan2020capturing}.

However, most numerical models are usually not easily differentiable. This major practical issue makes the optimization of ML components with online strategies difficult. For performance and historic reasons, legacy codes used in production are written in low-level programming languages such as Fortran or C. This is the case in many physics communities including the climate modeling community, which uses Earth System models for climate projections. These hand-optimized numerical models are also relatively large codebases ($\sim$ hundreds of thousands of lines of codes) and would thus require lots of effort to be re-implemented in a modern AD language. Generating adjoints automatically using code analysis libraries is possible, but practically complicated to deploy and lead to suboptimal codes (in particular in terms of memory footprint). In practice, online learning strategies are thus not easily amenable with production scale codes.
Different approaches that includes the temporal aspect provided by online learning have also been proposed and rely on gradient-free optimization algorithms. While gradient-free optimization is a large subfield, we identify two large families that have recently seen many applications to training large ML models in the context of hybrid physical models. First, methods adapted from data assimilation (DA) \cite{pawar2021data, huang2022efficient} and inverse problems as for instance the ensemble Kalman inversion \cite{iglesias2013ensemble, kovachki2019ensemble, lopez2022training}. More recently, many studies applying gradient-free policy search in reinforcement learning have been explored in particular for turbulence modeling \cite{kim2022deep, kurz2023deep, novati2021automating, bae2022scientific}. Promising avenues based on hybrid approaches using reservoid computing \cite{wikner2020combining} have also been applied to atmospheric modeling \cite{arcomano2022hybrid,arcomano2023hybrid} Overall, it is not clear which strategy performs best or should be preferred in practical situations.

Alternatively, we here propose to leverage neural emulation as a strategy to perform online supervised learning while avoiding the need for a differentiable solver. The use of neural emulators also connects with some algorithms in the field of simulation based inference which use neural networks to solve inverse problems \cite{cranmer2020frontier} and the field of DA for model error correction \cite{nonnenmacher2021deep,bocquet2023surrogate}. Here, the neural emulator aims at approximating the numerical solver in the model state space, i.e., at coarse resolution. We can rephrase the neural emulation as an approximation of the temporal evolution of some PDE.
We describe in this paper a generic algorithm which allows to train jointly a ML-based SGS parametrization for a non-differentiable solver with an online strategy and a neural emulator of the same solver. To overcome bias separation between the two components, we introduce a two-step procedure and a compensated loss function that isolates the different training objectives. The algorithm is implemented and demonstrated to produce stable models on a challenging subgrid parametrization problem. We also show that we are able to recover most of the benefits of online strategies without having to compute the gradient of the coarse-resolution solver.

The paper is structured as follows: in Section \ref{sec:problem-statement}, we outline the general issue being addressed and its particular application to SGS modeling. Section \ref{sec:online-emulation} describes the two-step algorithm that allows training a SGS parametrization with online strategy using a differentiable neural emulator of the target solver. In Section \ref{sec:demonstration}, we demonstrate the proposed methodology on a simple chaotic Lorenz-96 system. Finally, the same generic methodology is applied in \ref{sec:application} to evaluate its potential on a two-dimensional quasi-geostrophic system for a set of instantaneous and statistical \textit{a priori} and \textit{a posteriori} metrics. The results are further discussed in Section \ref{sec:discussion}.

\section{Learning strategies for model correction}
\label{sec:problem-statement}
As explained in the introduction, training a model online requires the ability to compute the gradient of the loss function with respect to model parameters, which involves solver calls and consequently also requires its gradient.
Let us define a typical regression task where one would seek to minimize a cost between some prediction $\hat{\mathbf{z}}_{\theta} = \mathcal{M}_{\theta}(\mathbf{y})$ and ground truth of the same input space $\mathbf{z}$, i.e.
\begin{equation}
  \theta^{\star} = \operatorname*{arg\,min}_\theta \mathcal{L}(\mathbf{z}, \hat{\mathbf{z}}_{\theta}).
  \label{eq:min}
\end{equation}
During the training process, the parameters of $\mathcal{M}$, $\theta$ are optimized in order to minimize $\mathcal{L}$. 
The minimization algorithm requires the computation of the gradient of the loss function with respect to the trainable parameters $\theta$,
\begin{equation}
  \frac{\partial \mathcal{L}}{\partial \theta}(\mathbf{z}, \hat{\mathbf{z}}_{\theta}) = \left( \frac{\partial \mathcal{M}_{\theta}(\mathbf{y})}{\partial \theta} \right)^{\mathsf{T}} \frac{\partial \mathcal{L}}{\partial \mathcal{M}_{\theta}}.
  \label{eq:minimization-differentiable}
\end{equation}
In practice, $\mathcal{M}$ could be a simple operator from which the analytical gradient would be computed beforehand and provided to the optimization process.
However, in differentiable solvers, we are interested in the temporal evolution of a system $E(\mathbf{y}): \mathbb{R}^{n} \rightarrow \mathbb{R}^{n}$ of a vector-valued quantity of interest $\mathbf{y}(t)$.
In a discretized formulation, the operator $E$ is typically defined by a sequence of operations,
\begin{equation}
  \mathbf{y}(t + \Delta t) = E_{p} \circ \cdots \circ E_{1}(\mathbf{y}(t)).
\end{equation}
Now, if $\mathcal{M} \equiv E$, the required partial derivative of the system is defined as a Jacobian in the minimization formulation \eqref{eq:minimization-differentiable},
\begin{equation}
  \frac{\partial E_{i}}{\partial \theta} = \begin{bmatrix}
						  \dfrac{\partial E_{i, 1}}{\partial \theta_{1}} & \cdots & \dfrac{\partial E_{i, 1}}{\partial \theta_{n}} \\
						  \vdots & \ddots & \vdots \\
						  \dfrac{\partial E_{i, m}}{\partial \theta_{1}} & \cdots & \dfrac{\partial E_{i, m}}{\partial \theta_{n}}
					       \end{bmatrix}
\end{equation}
for a single step of the temporal evolution operator $E$. 
Composing the partial derivatives of the sequences of operators in $E$ can be done by applying the chain rule,
\begin{equation}
  \frac{\partial E}{\partial \theta} = \frac{\partial (E_{p} \circ \cdots \circ E_{1})}{\partial \theta} = \frac{\partial E_{p}}{\partial E_{p - 1}} \cdots \frac{\partial E_{2}}{\partial E_{1}} \frac{\partial E_{1}}{\partial \theta}
  \label{eq:jacobians-chain}
\end{equation}
The gradient of $E$ can quickly become difficult to maintain by hand.
Note that when training NNs, it is required to have a scalar-valued loss function $\mathcal{L}$, which means that its derivative is a gradient instead of a Jacobian.
The way partial derivatives are computed is called reverse-mode differentiation \cite{baydin2018automatic}, which here consists of a series of vector-matrix multiplications starting from
\begin{equation}
  \left( \frac{\partial \mathcal{L}}{\partial E} \right)^{\mathsf{T}} \frac{\partial E_{p}}{\partial \theta} \in \mathbb{R}^{n}.
\end{equation}
Solving \eqref{eq:jacobians-chain} can be done by implementing a solver leveraging AD languages or libraries.
However, as pointed out in the introduction, this is a challenging task for large-scale solvers such as GCMs that would requires a lot of development work and might lead to the vanishing gradient problem \cite{hochreiter1998vanishing} due to many function compositions.

\subsection{The specific SGS modeling problem}
Here, we are interested in modeling small-scale quantities arising in differential equations restricting the grid resolution and thus the performance of the simulations \cite{sagaut2006large}. We assume an underlying differential equation involving the time evolution of variables $\mathbf{y}(t)$ to be known and defined by an operator $f(\mathbf{y})$. 
The aim is to solve an equivalent approximation for coarse variables $\bar{\mathbf{y}}(t)$ such that
\begin{align}
  \begin{cases}
  \dfrac{\partial \mathbf{y}}{\partial t} = f(\mathbf{y}), \hspace{5mm} &\mathbf{y} \in \Omega \\[1.5ex]
  \dfrac{\partial \bar{\mathbf{y}}}{\partial t} = g(\bar{\mathbf{y}}) + \mathcal{M}(\bar{\mathbf{y}}), \hspace{5mm} &\bar{\mathbf{y}} \in \bar{\Omega} \\[1.5ex]
  \mathcal{T}(\mathbf{y}) = \bar{\mathbf{y}}
  \end{cases}
  \label{eq:learning-coarse}
\end{align}
where $\bar{\Omega} \subset \Omega$, $g$ is a coarse-resolution operator, $\mathcal{M}$ is a SGS model and $\mathcal{T}$ is a projection that maps variables to coarse resolution.
The objective is formulated as an inverse problem where operator $\mathcal{M}$ has to be determined such that the evolution of the coarse variables match the projection $\mathcal{T}(\mathbf{y})$ of the variables $\mathbf{y}$. In most situations, we have $f = g$ with variables existing on different spaces or dimensionalities. Note that this can be applied to any type of PDE without any loss of generality.
Within a learning framework, one states the identification of a general SGS term
\begin{equation}
  \tau(\mathbf{y}) = \mathcal{T}(f(\mathbf{y})) - g(\mathcal{T}(\mathbf{y})) \approx \mathcal{M}(\bar{\mathbf{y}};\theta_{\mathcal{M}})
  \label{eq:sgs-term}
\end{equation}
where $\theta_{\mathcal{M}}$ are trainable model parameters. Under the assumption that $\mathcal{T}$ commutes with partial derivatives, the classical approach comes to train a model $\mathcal{M}(\bar{\mathbf{y}};\theta_{\mathcal{M}})$ as a functional approximation of the missing term $\tau(\mathbf{y})$. The classical strategy to optimize $\mathcal{M}$ is to generate a dataset $\mathbb{D} = \{ \bar{\mathbf{y}} \} \rightarrow \{ \tau(\mathbf{y}) \}$ that maps input states $\bar{\mathbf{y}}$ to the SGS term $\tau(\mathbf{y})$ beforehand and then finding optimal parameters $\theta$ offline, i.e., without any knowledge of the dynamics $\partial \bar{\mathbf{y}} / \partial t$ during training. Here, let us refer to an offline objective $J_{\mathrm{off}}$, the corresponding minimization problem \eqref{eq:min} writes
\begin{equation}
    \operatorname*{arg\,min}_{\theta} J_{\mathrm{off}} \equiv \operatorname*{arg\,min}_{\theta} \mathcal{L}_{\mathrm{off}}\left( \tau(\mathbf{y}), \mathcal{M}_{\theta}(\bar{\mathbf{y}}) \right)
    \label{eq:min-off}
\end{equation}
where $\mathcal{L}_{\mathrm{off}}$ is an instantaneous scalar-valued loss function $\mathcal{L}_{\mathrm{off}}$ $: (\mathbb{R}^{d}, \mathbb{R}^{d}) \rightarrow \mathbb{R}$ that penalizes the NN prediction from the expected missing term. In this context, $\mathcal{M}$ is defined as a NN implemented in a standard deep learning framework such as PyTorch or TensorFlow. 
This allows for the computation of the gradient of the offline objective, i.e., injecting \eqref{eq:min-off} in \eqref{eq:minimization-differentiable},
\begin{equation}
    \frac{\partial J_{\mathrm{off}}}{\partial \theta} = \left( \frac{\partial \mathcal{L}_{\mathrm{off}}}{\partial \mathcal{M}_{\theta}} \right)^{\mathsf{T}} \frac{\partial \mathcal{M}_{\theta}(\bar{\mathbf{y}})}{\partial \theta}.
\end{equation}
Using this formulation, the problem is simple to setup since both terms on the right-hand side are automatically determined using AD capabilities. However, while the offline approach is efficient to accurately reproduce the statistical properties of the missing term $\tau(\mathbf{y})$, it has been shown to lead to unstable simulations in non-linear, chaotic dynamical systems such as those found in turbulence modeling \cite{maulik2019subgrid}. Improving the stability of the SGS model can be done by increasing the size of the dataset, in particular by reducing the inter-sample correlation \cite{guan2022stable} or adding physics-based terms in the loss function \cite{guan2023learning}. As pointed out in \cite{frezat2022posteriori}, the offline training approach may not be able to optimize a model based on relevant metrics in the context of a dynamical system. An equivalent training approach tailored to this task, called online, has already been explored successfully in turbulence modeling \cite{um2020solver,holl2020learning,macart2021embedded,kochkov2021machine}. Formally, \citeA{frezat2022posteriori} described the minimization problem \eqref{eq:min} of the online loss $\mathcal{L}_{\mathrm{on}}$ where
\begin{equation}
    \operatorname*{arg\,min}_{\theta} J_{\mathrm{on}} \equiv \operatorname*{arg\,min}_{\theta} \mathcal{L}_{\mathrm{on}}\left( \{\mathbf{y}(t)\}_{t \in [t_{0},t_{1}]}, \{\Phi_{\theta}^{t}(\bar{\mathbf{y}}(t_{0}))\}_{t \in [t_{0},t_{1}]} \right)
    \label{eq:min-on}
\end{equation}
where $\Phi$ is the flow operator that advances the coarse variables in time using the currently training model $\mathcal{M}$,
\begin{equation}
    \Phi_{\theta}^{t}(\bar{\mathbf{y}}(t_{0})) = \bar{\mathbf{y}}(t_{0}) + \int_{t_{0}}^{t} \left[ g(\bar{\mathbf{y}}(t^{\prime})) + \mathcal{M}_{\theta}(\bar{\mathbf{y}}(t^{\prime})) \right] \, \mathrm{d}t^{\prime}
    \label{eq:flow}
\end{equation}
and $\mathcal{L}_{\mathrm{on}}$ is a temporal scalar-valued loss function $\mathcal{L}_{\mathrm{on}}$ $: (\mathbb{R}^{d \times N_s}, \mathbb{R}^{d \times N_s}) \rightarrow \mathbb{R}$ defined on $N_s$ steps that discretizes the temporal interval $[t_{0}, t]$. Let us expand the gradient of the online objective $J_{\mathrm{on}}$ from \eqref{eq:minimization-differentiable} at final time $t$,
\begin{equation}
    \frac{\partial J_{\mathrm{on}}}{\partial \theta} = \left( \frac{\partial \mathcal{L}_{\mathrm{on}}}{\partial \Phi_{\theta}^{t}} \right)^{\mathsf{T}} \frac{\partial \Phi_{\theta}^{t}}{\partial \theta}(\bar{\mathbf{y}}(t_{0})) = \left( \frac{\partial \mathcal{L}_{\mathrm{on}}}{\partial \Phi_{\theta}^{t}} \right)^{\mathsf{T}} \left[ \int_{t_{0}}^{t} \left( \frac{\partial g(\bar{\mathbf{y}}(t^{\prime}))}{\partial \theta} + \frac{\partial \mathcal{M}_{\theta}(\bar{\mathbf{y}}(t^{\prime}))}{\partial \theta} \right) \, \mathrm{d}t^{\prime} \right].
\end{equation}
This process requires the ability to compute the partial derivatives of the coarse-resolution solver $g$ with respect to the trained model parameters $\theta$, i.e. $\partial g / \partial \theta$. Here we are interested in determining this quantity such that $\tau$ can be learnt online without re-implementing $g$ using an AD framework.

\section{Online learning from differentiable emulation}
\label{sec:online-emulation}
Building on \cite{nonnenmacher2021deep}, we propose a learning scheme that is able to train a differentiable emulator of $g$ that can be used temporarily to train a SGS term with gradient information. Let us introduce a differentiable emulator $\mathcal{E}$ of $g$
\begin{equation}
    \mathcal{E}(\bar{\mathbf{y}}) \approx g(\bar{\mathbf{y}}), \hspace{1mm} \text{such that } \frac{\partial \mathcal{E}}{\partial \theta} \text{ is known}.
\end{equation}
This differentiability condition is automatically satisfied if $\mathcal{E}$ is implemented using AD capabilities. It is possible to define $\mathcal{E}$ as a trainable neural emulator, which involves training its parameters $\vartheta$ jointly with the SGS model parameters $\theta$. Substituting into \eqref{eq:min-on},
\begin{equation}
    \operatorname*{arg\,min}_{\vartheta, \theta} J_{\mathrm{on}}^{\mathrm{coupled}} \equiv \operatorname*{arg\,min}_{\vartheta, \theta} \mathcal{L}_{\mathrm{on}}\left( \{\mathbf{y}(t)\}_{t \in [t_{0},t_{1}]}, \{\bar{\mathbf{y}}(t_{0}) + \int_{t_{0}}^{t} \left[ \mathcal{E}_{\vartheta}(\bar{\mathbf{y}}(t^{\prime})) + \mathcal{M}_{\theta}(\bar{\mathbf{y}}(t^{\prime})) \right] \, \mathrm{d}t^{\prime}\}_{t \in [t_{0},t_{1}]} \right).
    \label{eq:coupled-min}
\end{equation}
We recall here that the final goal of this learning scheme is to provide a stable SGS model $\mathcal{M}$ that will be used for forward simulations with the coarse-resolution solver $g$. However, with the coupled formulation \eqref{eq:coupled-min}, the model $\mathcal{M}$ is not explicitly restricted to represent SGS dynamics only. Indeed the learning scheme jointly optimizes $\vartheta$ and $\theta$ such that $\mathcal{E} + \mathcal{M}$ approximates the solver $f$. There is therefore no guarantee that $\theta$ will be optimal for $\mathcal{M}$ to represent the SGS term.

\subsection{Sequential neural emulation-SGS learning}
It is possible to penalize both the neural emulator and the SGS model by composing the loss function with weight $\lambda \in [0, 1] \subset \mathbb{R}$ such that $\mathcal{L}_{\mathrm{on}}^{\mathrm{composed}} \equiv \lambda \mathcal{L}_{\mathrm{on}}^{\mathcal{E}}(\mathcal{E}_{\vartheta}(\bar{\mathbf{y}})) + (1 - \lambda) \mathcal{L}_{\mathrm{on}}^{\mathcal{M}}(\mathcal{M}_{\theta}(\bar{\mathbf{y}}))$ where $\mathcal{L}_{\mathrm{on}}^{\mathcal{E}}$ and $\mathcal{L}_{\mathrm{on}}^{\mathcal{M}}$ targets the neural emulator and SGS model, respectively. It might be possible to use filtering operations to identify their effect on high-resolution reference data, but their respective definitions may not be simple to implement and might not include inter-scale transfers.
Instead, we propose a two-step online formulation that trains $\mathcal{E}$ and $\mathcal{M}$ sequentially,
\begin{align}
    &\textbf{Step 1 (neural emulator):} \nonumber\\
    &\operatorname*{arg\,min}_{\Theta} \ell_{\mathrm{on}}^{\mathcal{E}}\left( \{\bar{\mathbf{y}}(t)\}_{t \in [t_{0},t_{1}]}, \{\bar{\mathbf{y}}(t_{0}) + \int_{t_{0}}^{t} \mathcal{E}(\bar{\mathbf{y}}(t^{\prime});\Theta)  \, \mathrm{d}t^{\prime}\}_{t \in [t_{0},t_{1}]} \right), \label{eq:dual-emu} \\
    &\textbf{Step 2 (subgrid-scale model):} \nonumber\\
    &\operatorname*{arg\,min}_{\theta} \ell_{\mathrm{on}}^{\mathcal{M}}\left( \{\mathbf{y}(t)\}_{t \in [t_{0},t_{1}]}, \{\bar{\mathbf{y}}(t_{0}) + \int_{t_{0}}^{t} \mathcal{E}(\bar{\mathbf{y}}(t^{\prime})) + \mathcal{M}(\bar{\mathbf{y}}(t^{\prime});\theta)  \, \mathrm{d}t^{\prime}\}_{t \in [t_{0},t_{1}]} \right). \label{eq:dual-sgs}
\end{align}
These two steps are formulated using an online learning strategy, but \eqref{eq:dual-emu} minimizes a loss for an integration of the coarse-resolution solver $g$ while \eqref{eq:dual-sgs} is equivalent to the classical online SGS formulation \eqref{eq:min-on} with $\mathcal{E} \approx g$.
This sequential training scheme does not require a differentiable version of the coarse-resolution solver $g$. Indeed, let us factor the modified flows $\varphi$ appearing on the right-hand side of \eqref{eq:dual-emu} and \eqref{eq:dual-sgs} respectively,
\begin{align}
    \varphi_{\Theta}^{t}(\bar{\mathbf{y}}(t_{0})) &= \bar{\mathbf{y}}(t_{0}) + \int_{t_{0}}^{t} \mathcal{E}(\bar{\mathbf{y}}(t^{\prime});\Theta) \, \mathrm{d}t^{\prime}, \\
    \varphi_{\theta}^{t}(\bar{\mathbf{y}}(t_{0})) &= \bar{\mathbf{y}}(t_{0}) + \int_{t_{0}}^{t} \mathcal{E}(\bar{\mathbf{y}}(t^{\prime})) + \mathcal{M}(\bar{\mathbf{y}}(t^{\prime});\theta)  \, \mathrm{d}t^{\prime}.
\end{align}
Then, expanding again the gradient of the online losses using \eqref{eq:minimization-differentiable} yields,
\begin{align}
    \left( \frac{\partial \ell_{\mathrm{on}}^{\mathcal{E}}}{\partial \varphi_{\Theta}^{t}}\right )^{\mathsf{T}} \frac{\partial \varphi_{\Theta}^{t}}{\partial \Theta}(\bar{\mathbf{y}}(t_{0})) &= \left( \frac{\partial \ell_{\mathrm{on}}^{\mathcal{E}}}{\partial \varphi_{\Theta}^{t}}\right )^{\mathsf{T}} \left( \int_{t_{0}}^{t} \frac{\partial \mathcal{E}(\bar{\mathbf{y}}(t^{\prime});\Theta)}{\partial \Theta} \, \mathrm{d}t^{\prime} \right), \\
    \left( \frac{\partial \ell_{\mathrm{on}}^{\mathcal{M}}}{\partial \varphi_{\theta}^{t}}\right )^{\mathsf{T}} \frac{\partial \varphi_{\theta}^{t}}{\partial \theta}(\bar{\mathbf{y}}(t_{0})) &= \left( \frac{\partial \ell_{\mathrm{on}}^{\mathcal{M}}}{\partial \varphi_{\theta}^{t}}\right )^{\mathsf{T}} \left( \int_{t_{0}}^{t} \frac{\partial \mathcal{E}(\bar{\mathbf{y}}(t^{\prime}))}{\partial \theta} + \frac{\partial \mathcal{M}(\bar{\mathbf{y}}(t^{\prime});\theta)}{\partial \theta} \, \mathrm{d}t^{\prime} \right)
\end{align}
with $\mathcal{E}$, $\mathcal{M}$ are arbitrary NNs implemented using AD and $\ell_{\mathrm{on}}^{\mathcal{E}}$, $\ell_{\mathrm{on}}^{\mathcal{M}}$ loss functions also defined within the ML framework, supporting AD by construction.

\subsection{Neural emulator error compensations}
\label{subsec:error-compensations}
In theory, our two-step learning scheme would be equivalent to the classical online SGS learning scheme with a differentiable solver if and only if $\mathcal{E} \equiv g$. In practice however, the neural emulator will only represent an approximation of the coarse-resolution solver, i.e., $\mathcal{E} \approx g$. The performance of the SGS training \eqref{eq:dual-sgs} thus strongly relies on the emulator's ability to reproduce the adjoint (see. \citeA{ouala2024online} for an example) of the coarse dynamics on a given time horizon $[t_{0}, t]$. The instantaneous emulation mismatch between the solver and emulator can be expressed as 
\begin{equation}
    \varepsilon(\bar{\mathbf{y}}) = g(\bar{\mathbf{y}}) - \mathcal{E}(\bar{\mathbf{y}}).
\end{equation}
In practice, the loss function \eqref{eq:dual-sgs} always acts on coarse resolution, i.e., it takes the projection $\mathcal{T}$ on data $\mathbf{y}(t)$. Substituting the projected SGS term \eqref{eq:sgs-term} on the left-hand side gives the ``exact" coarse flow,
\begin{equation}
    \mathcal{T}\left( \mathbf{y}(t) \right) = \mathcal{T}\left( \mathbf{y}(t_{0}) + \int_{t_{0}}^{t} f(\mathbf{y}(t^{\prime})) \, \mathrm{d}t^{\prime} \right) = \bar{\mathbf{y}}(t_{0}) + \int_{t_{0}}^{t} g(\bar{\mathbf{y}}(t^{\prime})) + \tau(\mathbf{y}(t^{\prime})) \, \mathrm{d}t^{\prime}
\end{equation}
Now, expanding the modified flow $\varphi_{\theta}^{t}(\bar{\mathbf{y}}(t_{0})$ with the emulation mismatch in the right-hand side of \eqref{eq:dual-sgs}, we see that if $\mathcal{M}$ approximates $\tau$, the flow is equivalent only if $\epsilon$ vanishes,
\begin{align}
    \varphi_{\theta}^{t}(\bar{\mathbf{y}}(t_{0}) = \bar{\mathbf{y}}(t_{0}) + \int_{t_{0}}^{t} g(\bar{\mathbf{y}}(t^{\prime})) - \varepsilon(\bar{\mathbf{y}}(t^{\prime})) + \mathcal{M}(\bar{\mathbf{y}}(t^{\prime}) ;\theta)  \, \mathrm{d}t^{\prime}
    \label{eq:expanded_flow}
\end{align}
It is clear here that the emulation mismatch $\varepsilon$ will be absorbed by the SGS model $\mathcal{M}$ during its training phase, since $\mathcal{E}$ is already fixed. In order to optimize $\mathcal{M}$ against $\tau$ using this standard online loss function, it is important that the emulation mismatch remains small.
Controlling this mismatch is challenging in particular due to the temporal integration. Recall that the loss function $\ell_{\mathrm{on}}^{\mathcal{M}}$ is defined on a sequence of states of the system $\bar{\mathbf{y}}(t)$ but never involves the missing SGS term $\tau$ that $\mathcal{M}$ is trying to learn. We instead propose to re-target the loss function $\ell_{\mathrm{on}}^{\mathcal{M}}$ so that it directly operates on the SGS terms for the considered time window,
\begin{equation}
    \ell_{\mathrm{on}}^{\mathrm{subgrid}}(\{\mathbf{y}(t)\}_{t \in [t_{0},t_{1}]}, \{\varphi_{\theta}^{t}(\bar{\mathbf{y}}(t_{0})\}_{t \in [t_{0},t_{1}]}) = \ell_{\mathrm{on}}^{\mathcal{M}}(\{\tau(\mathbf{y}(t))\}_{t \in [t_{0},t_{1}]}, \{\mathcal{M}(\varphi_{\theta}^{t}(\bar{\mathbf{y}}(t_{0}))\}_{t \in [t_{0},t_{1}]})
\end{equation}
In practice, this loss is still online because it involves temporal integration through the flow operator. Expanding \eqref{eq:expanded_flow} at final trajectory time $t$, we observe that this loss now acts as a model composition of the bias emulation mismatch $\varepsilon$,
\begin{equation}
    \mathcal{M}(\varphi_{\theta}^{t}(\bar{\mathbf{y}}(t_{0})) = \mathcal{M}\left( g(\bar{\mathbf{y}}(t)) - \varepsilon(\bar{\mathbf{y}}(t)) + \mathcal{M}(\bar{\mathbf{y}}(t)) \right)
\end{equation}
Here, the first benefit is that model $\mathcal{M}$ is now trained to be invariant to the emulator mismatch $\varepsilon$ with respect to the SGS term $\tau$. The second benefit comes from the small-scale target of the loss function, absorbing the large-scale emulator mismatch $\varepsilon$ into a small-scale error. From experiments, we observed that penalizing the SGS terms in the loss function largely reduces the large-scale ambiguities from the emulation mismatch $\varepsilon$ while maintaining the online behavior of the learning strategy, since these quantities are still evaluated over a temporal horizon $[t_{0}, t]$. From now, we will refer to the original loss $\ell_{\mathrm{on}}^\text{state} \equiv \ell_{\mathrm{on}}^{\mathcal{M}}$ as ``state"-based online loss and the re-targeting loss $\ell_\text{on}^\text{subgrid}$ as ``subgrid"-based online loss. Experiments based on emulators in the next section will be designed on the subgrid loss, since training with the original state loss did not lead to stable models. The discretized version of the two-step learning scheme, suitable for numerical applications (detailed below), is outlined in Algorithm \ref{alg:twostep}.
\draftfalse
\begin{algorithm}[tb]
\caption{Two-step online SGS (model) learning scheme}\label{alg:twostep}
\begin{algorithmic}
\renewcommand{\algorithmicrequire}{\textbf{input:}}
\renewcommand{\algorithmicensure}{\textbf{output:}}
\renewcommand{\algorithmiccomment}[1]{\hfill \texttt{//} \textsc{#1}}
\Require{coarse-grained dataset $\{\bar{\mathbf{y}}(t)\}$ computed from $g$} 
\Require{solver dataset $\{\mathbf{y}(t), \tau(\mathbf{y}(t))\}$ computed from $f$} 
\Require{emulator architecture and parameters $\mathcal{E}(\Theta \sim \mathcal{N})$} 
\Require{model architecture and parameters $\mathcal{M}(\theta \sim \mathcal{N})$} 

\Comment{Step 1: neural emulator, Eq. \eqref{eq:dual-emu}}
\While{emulator parameters not converged}
    \State randomly select a temporal interval: $t = [t_{0}, t_{1}]$
    \State initialize from coarse states: $\bar{\mathbf{y}}_{\mathcal{E}}(t_{0}) = \bar{\mathbf{y}}(t_{0})$
    \State integrate the emulator using ODE solver: $\{\bar{\mathbf{y}}_{\mathcal{E}}(t)\}_{t \in [t_{0},t_{1}]} = \Call{ode}{\mathcal{E}(\bar{\mathbf{y}}_{\mathcal{E}}(t_{0});\Theta)}$
    \State optimize emulator parameters on ``state" loss: $\ell_{\mathrm{on}}^{\mathcal{E}}(\{\bar{\mathbf{y}}(t)\}_{t \in [t_{0},t_{1}]}, \{\bar{\mathbf{y}}_{\mathcal{E}}(t)\}_{t \in [t_{0},t_{1}]})$
\EndWhile

\Comment{Step 2: subgrid-scale model, $\Theta$ fixed, Eq. \eqref{eq:dual-sgs}}
\While{model parameters not converged}
    \State randomly select a temporal interval: $t = [t_{0}, t_{1}]$
    \State initialize from projected states: $\hat{\mathbf{y}}(t_{0}) = \mathcal{T}(\mathbf{y}(t_{0}))$
    \State integrate system using ODE solver: $\{\hat{\mathbf{y}}(t)\}_{t \in [t_{0},t_{1}]} = \Call{ode}{\mathcal{E}(\hat{\mathbf{y}}(t_{0})) + \mathcal{M}(\hat{\mathbf{y}}(t_{0});\theta)}$
    \State optimize model parameters on ``subgrid" loss:
    $\ell_{\mathrm{on}}^{\mathrm{subgrid}}(\{\tau(\mathbf{y})(t)\}_{t \in [t_{0},t_{1}]}, \{\hat{\mathbf{y}}(t)\}_{t \in [t_{0},t_{1}]})$
\EndWhile
\Statex
\Ensure{optimized model parameters $\mathcal{M}(\theta)$} 
\end{algorithmic}
\end{algorithm}
\drafttrue

\section{Demonstration: two-timescales Lorenz-96 system}\noindent
\label{sec:demonstration}
In \cite{lorenz1996predictability}, Lorenz described a two-time scale dynamical system that mimics the non-linear dynamics of the extratropical atmosphere. This system, often referred to as the Lorenz-96 (L96) system, is a simplified representation of multiscale interactions and non-linear advection found in climate models. The L96 system is defined by two sets of equations coupling two sets of variables $X_k$ and $Y_{j,k}$ that evolves over slow and fast timescales, respectively,
\begin{align}
    \frac{d}{dt} X_k &= -X_{k - 1} (X_{k - 2} - X_{k + 1}) - X_k + F - \underbrace{\frac{hc}{b} \sum_{j = 1}^J Y_{j,k}}_{\tau_k} \label{eq:l96_slow} \\
    \frac{d}{dt} Y_{j,k} &= -cb Y_{j + 1, k} (Y_{j + 2, k} - Y_{j - 1, k}) - c Y_{j,k} + \frac{hc}{b} X_k \label{eq:l96_fast}
\end{align}
where $X_k \equiv \overline{\mathbf{y}}(t)$ denotes the slow (or coarse) $K$ variables, $Y_{j,k}$ denotes the fast $K \times J$ variables and $\tau_k$ is the coupling (or SGS) term since it involves fast variables $Y_{j,k}$ that are unknown when only slow variables $X_k$ are solved. In particular, we say here that the solver $f$ is equivalent to the two-timescales system \eqref{eq:l96_slow}-\eqref{eq:l96_fast}, where both $X_k$ and $Y_{j,k}$ are advanced in time. The coarse solver $g$ only solves for \eqref{eq:l96_slow} and must be provided with a parametrization of the SGS term $\tau_k$, and we are thus looking for a model $\mathcal{M}(X_k) \approx \tau_k$. The system is controlled by parameters $b$, $c$ and $h$ that determines the magnitude, fluctuation and strength of the fast variables compared to the slow ones. Finally, the slow time-scale equation is forced by $F$, a scalar that determines the chaotic behavior of the system. Note that the L96 has already been studied extensively in the context of DA and parametrization, including with online learning approaches (see \citeA{balwada2024learning} for a review).

\begin{table}
  \caption{Simulation (top) and learning (bottom) parameters used for the L96 demonstration.}
  \centering
  \begin{tabular}{l l c c}
  \\
  \toprule
  \textbf{Name} & \textbf{Symbol} & \textbf{Value} \\
  \midrule
  \textit{Simulation parameters} & & \\
  \midrule
   & $b$ & $10$\\
   & $c$ & $1$ \\
   & $h$ & $1$ \\
  Forcing & $F$ & $18$ \\
  Slow variables & $K$ & $8$ \\
  Fast variables & $J$ & $20$ \\
  Time step & $\delta t$ & $10^{-3}$ \\
  \midrule
  \textit{Learning parameters} & & $\mathcal{E}$ & $\mathcal{M}$\\
  \midrule
  Trainable parameters & & $26,056$ & $1,096$ \\
  Time steps per trajectory & $N_t$ & \multicolumn{2}{c}{$30$} \\
  Training trajectories & $N_\text{traj}$ & \multicolumn{2}{c}{$25$} \\
  \bottomrule
  \end{tabular}
  \label{table:l96-params}
\end{table}

\begin{figure}[tb]
  \centering
  \noindent\includegraphics[width=.99\textwidth]{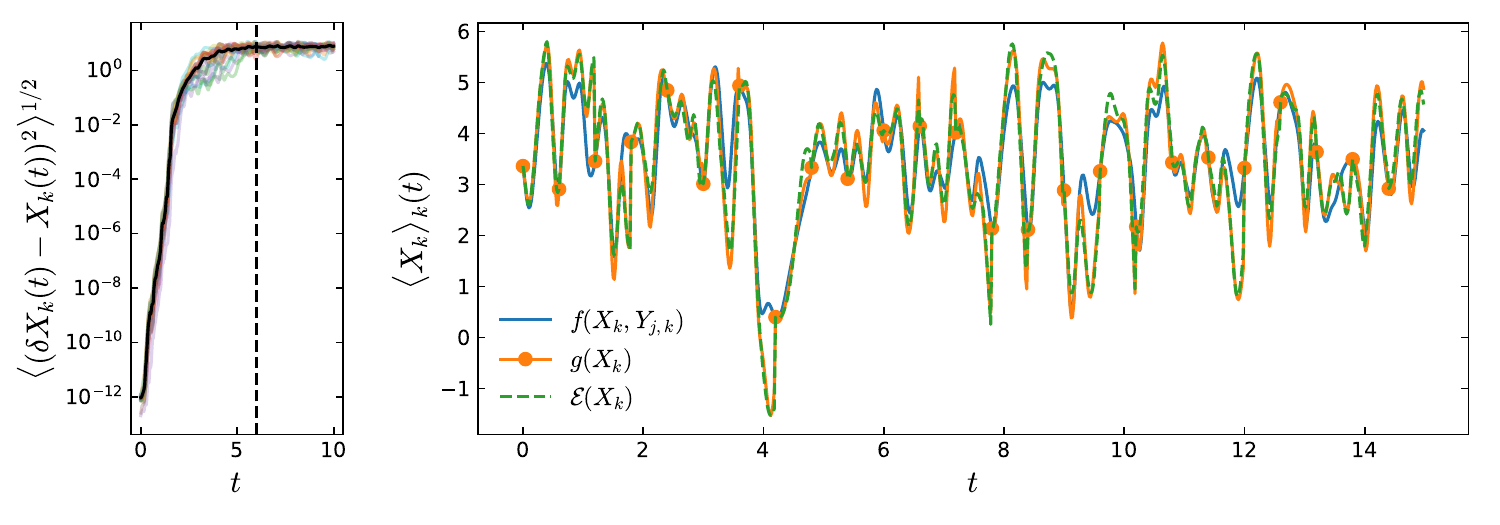}
  \caption{Evolution of the deviation (based on the root mean squared error) between reference trajectory $X_k$ and the perturbed ensemble $\delta X_k$ for the L96 system (left). The dashed lines shows the approximate decorrelation time $t_c$. Evolution of the mean state $\langle X_k \rangle_k$ throughout the 25 trajectories (right) for the reference solver $f$ (blue), coarse solver $g$ with restarts at each sub-trajectory (indicated by a circle), and neural emulator $\mathcal{E}$ trained to reproduce the dynamics of $g$.}
  \label{fig:l96-demo}
\end{figure}

The ODEs \eqref{eq:l96_slow}-\eqref{eq:l96_fast} are advanced in time using a classical order 4 Runge Kutta stepper, with control parameters $b$, $c$ and $h$ outlined in Table \ref{table:l96-params} along with $K = 8$ and $J = 20$ slow and fast variables, respectively. Because the L96 system is deterministically chaotic, we estimate the length of the training data trajectories as a small subset (10\%) of the decorrelation time of the system. Let us take a reference trajectory initialized with random Gaussian conditions $X_k(t = 0) \sim \mathcal{N}(0, b^2)$, $Y_{j,k}(t = 0) \sim \mathcal{N}(0, 1)$. The evolution of 25 perturbed trajectories such that $\delta X_k(t = 0) = X_k(t = 0) + 10^{-12} X_\text{pert}$ where $X_\text{pert} \sim \mathcal{N}(0, 1)$ gives us an estimate of the decorrelation time $t_c \approx 6$ (see Fig. \ref{fig:l96-demo}, left). With a time step $\delta t = 10^{-3}$, decorrelation time is reached in about $t_c \times 10^{3}$ steps using solver $f$. Note that with the coarse solver $g$, we can use larger time steps proportional to the number of fast variables, i.e., $\overline{\delta t} = J \delta t = 2 \times 10^{-2}$, which gives us trajectories of $N_t = 0.1 \times t_c \times \overline{\delta t}^{-1} = 30$ time steps. 

We show in Fig. \ref{fig:l96-demo} (right, blue line) the $N_\text{traj} = 25$ trajectories of coarse variables $X_k$ integrated using solver $f$ such that $\{ X_k(t) \}_{t \in [t_0,t_1]}$ where $t_1 - t_0 = 30 \overline{\delta t}$ . These states, along with the coupling term $\tau_k$ define the SGS training dataset for the Step 2 of Algorithm \ref{alg:twostep}.

For the coarse-grained dataset used in the neural emulator training (Step 1 of Algorithm \ref{alg:twostep}), states $X_k \equiv \hat{\mathbf{y}}(t)$ are generated by integrating the coarse solver $g$ restarting every $N_t$ steps from a reference state computed from $f$. Fig. \ref{fig:l96-demo} (right, orange line) shows how these states gradually diverge from the reference ones computed with $f$, before re-aligning with the reference (shown by a circle).

The NN architectures for the L96 system are based on multilayer perceptrons, i.e., multiple fully-connected blocks with ReLU non-linear activations. The neural emulator $\mathcal{E}$ contains 3 residual blocks operating on a 64-features latent space, while the SGS model $\mathcal{M}$ consists of 3 feed-forward blocks with a 16-features latent space, both with $K = 8$ input and output features (see Table \ref{table:l96-params} for a summary and number of trainable parameters). Note that there is a major difference in complexity between the neural emulator and the SGS model -- indeed, the neural emulator will only be used for training the SGS model, while the SGS model has to be computationally efficient for inference on long trajectories. The neural emulator is trained using a mean squared error (MSE) loss function on the coarse variables obtained from $g$,
\begin{equation}
    \ell_\text{on}^\mathcal{E}(\{X_k(t)\}_{t \in [t_0,t_1]}, \{X_k^\mathcal{E}(t)\}_{t \in [t_0,t_1]}) = \frac{1}{N_t} \frac{1}{K} \sum_{t \in [t_0,t_1]} \sum_{k = 1}^K \left( X_k^\mathcal{E}(t) - X_k(t) \right)^2
\end{equation}
where $X_k^\mathcal{E}$ is the coarse state predicted by the emulator. Fig. \ref{fig:l96-demo} (right, green dashed line) shows that the neural emulator $\mathcal{E}$ is able to closely match the trajectory of the coarse solver $g$ after 20,000 epochs of training with the AdamW optimizer.

\begin{figure}[tb]
  \noindent\includegraphics[width=.9\textwidth]{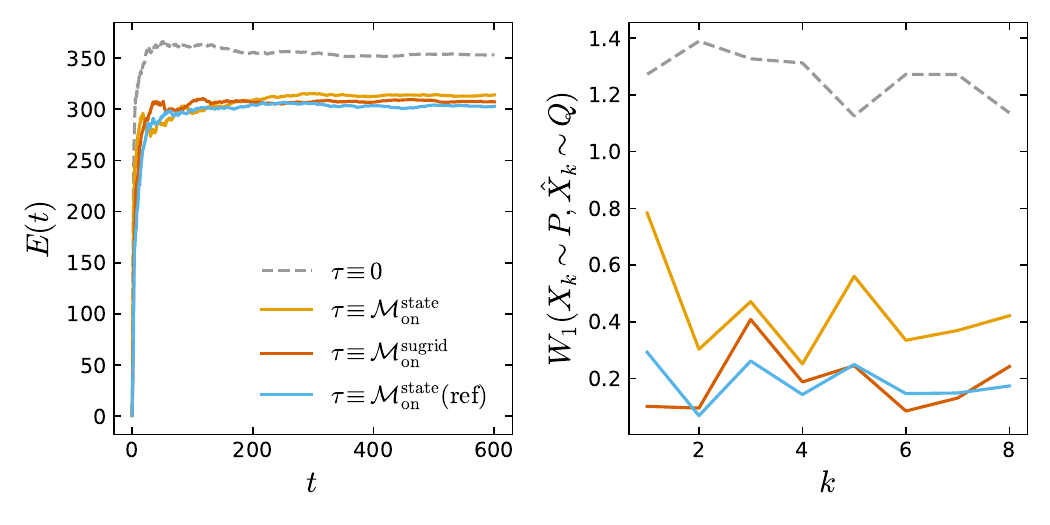}
  \caption{State-averaged cumulative error $E(t)$ (left) and Wasserstein distance $W_1(X_k \sim P, \hat{X}_k \sim Q)$ (right) for 100 decorrelation times $t_c$ (or 30,000 iterations with coarse solver $g$) for the L96 system.}
  \label{fig:l96-metrics}
\end{figure}

We now train the reference SGS model $\tau \equiv \mathcal{M}_\text{on}^\text{state} (\text{ref})$ with the differentiable solver $g$, and two models $\tau \equiv \mathcal{M}_\text{on}^\text{state}$, $\tau \equiv \mathcal{M}_\text{on}^\text{subgrid}$ using the emulator $\mathcal{E}$ with the state and subgrid losses, respectively. Here, the losses are defined as MSEs on the coarse variables obtained from $f$,
\begin{align}
    \ell_\text{on}^\text{state}(\{X_k(t)\}_{t \in [t_0,t_1]}, \{\hat{X}_k(t)\}_{t \in [t_0,t_1]}) &= \frac{1}{N_t} \frac{1}{K} \sum_{t \in [t_0,t_1]} \sum_{k = 1}^K \left( \hat{X}_k(t) - X_k(t) \right)^2, \\
    \ell_\text{on}^\text{subgrid}(\{\tau_k(t)\}_{t \in [t_0,t_1]}, \{\hat{X}_k(t)\}_{t \in [t_0,t_1]}) &= \frac{1}{N_t} \frac{1}{K} \sum_{t \in [t_0,t_1]} \sum_{k = 1}^K \left( \mathcal{M}_\text{on}^\text{subgrid}(\hat{X}_k(t); \theta) - \tau_k(t) \right)^2,
\end{align}
where $\hat{X}_k(t)$ is the state predicted by the trained model. After 5,000 epochs with the AdamW optimizer, these models are evaluated for 100 decorrelation times, which corresponds to 600,000 time steps with solver $f$ and 30,000 time steps with coarse solver $g$. Since the full dataset contains 25 trajectories, each taking 10\% of the decorrelation time, the evaluation then runs $40\times$ longer than the dataset trajectories horizon. 

The performance of the different models is evaluated on two diagnostics described in \cite{balwada2024learning}. The first one is a cumulative root-squared error metric on the model evolution, given by
\begin{equation}
    E(t) = \frac{1}{t} \int_0^t \sqrt{\left( \hat{X}_k(t') - X_k(t') \right)^2} \, \text{d}t'.
\end{equation}
We can see in Fig. \ref{fig:l96-metrics} (left) that trained models are able to perform better than the single-timescale simulation $\tau \equiv 0$, but the model trained with the state loss $\tau \equiv \mathcal{M}_\text{on}^\text{state}$ has a slightly larger error than the reference model and the one trained on the subgrid loss, which can be attributed to the imperfect nature of the states predicted by the emulator $\mathcal{E}$. The second metric measures the difference in empirical distributions based on the first Wasserstein distance
\begin{equation}
    W_1(X_k \sim P, \hat{X}_k \sim Q) = \inf_{\pi \in \Gamma(P, Q)} \int_{\mathbb{R} \times \mathbb{R}} |p - q| \, \text{d} \pi (p, q), \label{eq:wasserstein}
\end{equation}
computed separately for each state $k$. In Fig. \ref{fig:l96-metrics} (right), we see that the accuracy of the reference model $\tau \equiv \mathcal{M}_\text{on}^\text{state} (\text{ref})$ is only matched by the emulator-based model trained with the subgrid loss $\tau \equiv \mathcal{M}_\text{on}^\text{subgrid}$.

We have shown that the approach previously applied on a single-timescale Lorenz-96 system in \cite{nonnenmacher2021deep} extends to moderately more complex systems. However, this demonstration already reveals some limitations in the emulator's ability to accurately to reproduce the dynamics --and consequently, the adjoint-- of the reference solver. In particular, the metrics reported in Fig. \ref{fig:l96-metrics} suggest that some improvement is possible by considering a different loss function when training for the parameterization.

\section{Application: quasi-geostrophic dynamics}\noindent
\label{sec:application}
In turbulent flows, the energy cascade drives energy from the large scales to the small scales until molecular viscous dissipation (forward-scatter), but the inverse transfer called backscatter in which energy is transferred from the small scales back to the large scales \cite{lesieur1996new} is also in play, particularly for geophysical flows.
This is explained by the relative dominance of the Coriolis force which creates vortical structures that appear two-dimensional.
Historically, developing SGS models that account for backscatter is a challenging task \cite{piomelli1991subgrid,schumann1995stochastic,liu2011modification}.
Indeed, an overprediction of backscatter that can not be compensated by eddy-viscosity will lead to an accumulation of small-scale energy causing simulations to become numerically unstable.
In two-dimensional flows, we observe a dual cascade composed of ``forward'' enstrophy and ``inverse'' energy, in a statistical sense.
As a consequence, a large number of SGS models have been proposed in particular for geophysical flows (see \citeA{danilov2019toward} for a review)
with well-documented configurations and performance metrics \cite{fox2008can}.
SGS modeling is also a key issue for the simulation of ocean and atmosphere dynamics because of the large range of motions involved \cite{jansen2015energy,juricke2019ocean,juricke2020ocean,frederiksen2012stochastic}.
As a case study framework, we consider barotropic QG flows. While providing an approximate yet representative system for rotating stratified flows found in the atmosphere and ocean dynamics, it involves relatively complex SGS features that make the learning problem non-trivial. As such, QG flows are regarded as an ideal playground to explore and assess the relevance of ML strategies for SGS models in geophysical turbulence. The governing equations of vorticity $\omega$ for the QG model are
\begin{align}
  \dfrac{\partial \omega}{\partial t} + \mathcal{N}(\omega, \psi) &= \nu \nabla^{2} \omega  - \mu \omega - \beta \partial_{x} \psi + \mathcal{F}_{\omega}, \label{eq:qg} \\
  \nabla^{2}\psi &= -\omega
  \label{eq:qg-inc}
\end{align}
where $\mathcal{N}(\omega, \psi) = \partial_{x} \psi \partial_{y}(\omega) - \partial_{y} \psi \partial_{x}(\omega)$ is the non-linear vorticity advection, $\nu$ is the viscosity, $\mu$ is a large-scale drag coefficient, $\beta$ is the Earth rotation vector approximation by beta-plane and $\mathcal{F}_{\omega}$ is an additional source term. Now, the derivation of the coarse-resolution system for QG dynamics follows the same procedure that is typically used for fluid dynamics turbulence modeling (also called large-eddy simulation). The projection operator
$\mathcal{T}$ at spatial coordinate $\mathbf{x}$ is here given as a discretization (or coarse-graining) $\mathcal{D} : \Omega \rightarrow \bar{\Omega}$ and the convolution of $\omega$ with a kernel function $G(\mathbf{x})$ \cite{leonard1975energy},
\begin{equation}
    \bar{\omega}(\mathbf{x}) = \mathcal{D} \left( \int G(\mathbf{x} - \mathbf{x}^{\prime})\omega(\mathbf{x}^{\prime}) \, \mathrm{d}\mathbf{x}^{\prime} \right). \label{eq:filtering}
\end{equation}
We can then derive the equations which govern the evolution of vorticity $\bar{\omega}$, 
\begin{equation}
  \dfrac{\partial \bar{\omega}}{\partial t} + \mathcal{N}(\bar{\omega}, \bar{\psi}) = \nu \nabla^{2} \bar{\omega}  - \mu \bar{\omega} - \beta \partial_{x} \bar{\psi} + \underbrace{\mathcal{N}(\bar{\omega}, \bar{\psi}) - \overline{\mathcal{N}(\omega, \psi)}}_{\tau_{\omega}} + \bar{\mathcal{F}}_{\omega} \label{eq:qg-coarse}
\end{equation}
where $\tau_{\omega}$ is the SGS term. In this context, $\tau_{\omega}$ can not be determined from the coarse variables because of the non-linear interactions of small-scale dynamics $\overline{\mathcal{N}(\omega, \psi)}$ and must thus be statistically modeled in order to close equation \eqref{eq:qg-coarse}.
\begin{figure}[tb]
  \centering
  \noindent\includegraphics[width=.99\textwidth]{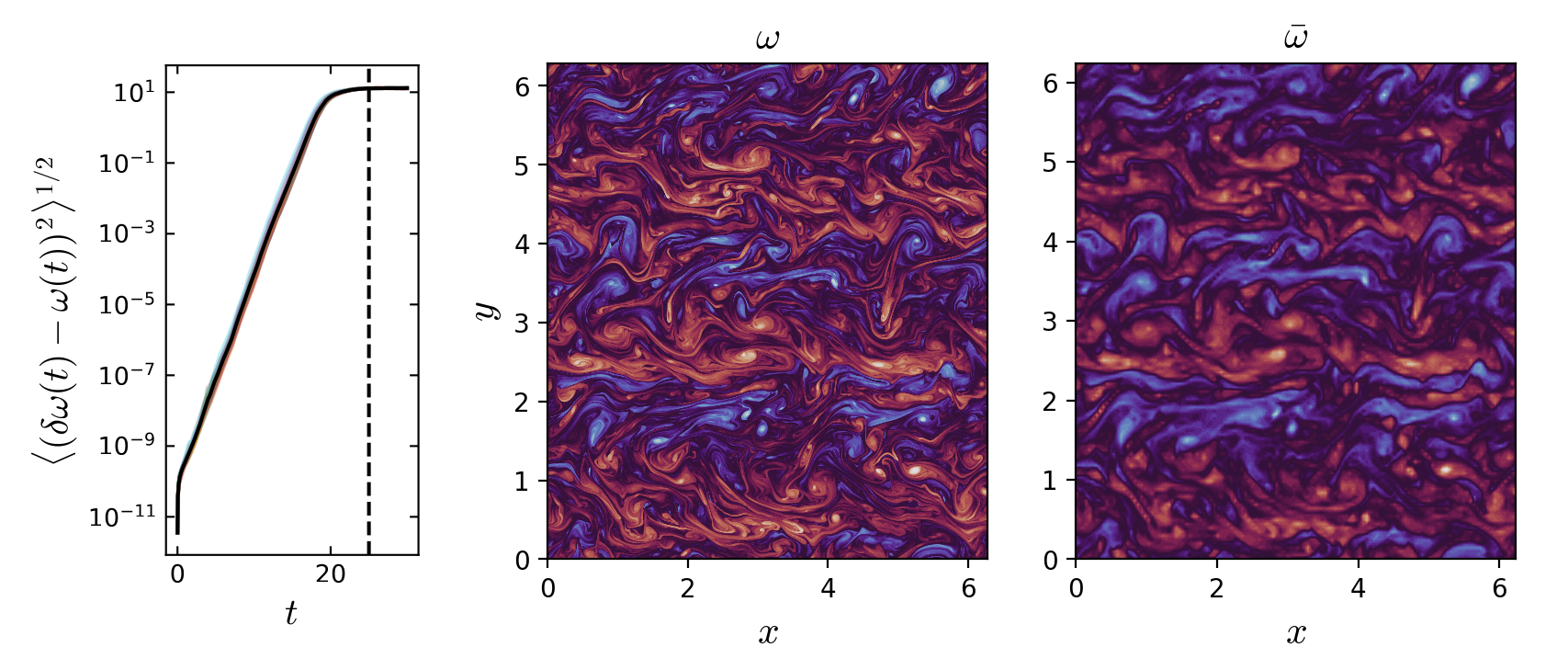}
  \caption{Evolution of the deviation (based on the root mean squared error) between reference trajectory $\omega(t)$ and the perturbed ensemble $\delta \omega(t)$ for the QG system (left). The dashed lines shows the approximate decorrelation time $t_c$. Vorticity field $\omega$ (center) and filtered vorticity field $\bar{\omega}$ (right) at the end of the spin-up.}
  \label{fig:qg-app}
\end{figure}

\subsection{Simulation setup}\noindent
In order to study the learning problem, we solve equations \eqref{eq:qg}-\eqref{eq:qg-inc} using a pseudospectral code with full 3/2 dealiasing \cite{canuto2007spectral} and a third-order implicit-explicit (IMEX) Runge Kutta time advancement scheme \cite{boscarino2013implicit}. The system is defined in a squared domain $\Omega, \bar{\Omega} = [0, 2\pi]^{2}$, with domain length $L = 2\pi$ discretized with a Fourier basis, i.e. double-periodic boundary conditions on $N$ and $\bar{N}$ grid points with uniform spacing and a grid ratio $\Delta^{\prime} = N/\bar{N} > 1$ between full and coarse solver resolutions. We extract the SGS term with spatial filtering \eqref{eq:filtering} using a spectral cut-off filter $G$ and sample non-residual quantities along the simulations, i.e. $\{\omega \rightarrow \tau_{\omega}\}$ is obtained from solver $f$ and $\{\bar{\omega}\}$ is obtained from coarse-resolution solver $g$, similarly to the methodology used for the L96 demonstration. 
We energize the system using a deterministic forcing \cite{guan2024online} at a given scale
\begin{equation}
    \mathcal{F}_{\omega}(x, y) =  \sigma\cos(k_f x) + \sigma\cos(k_f y),
\end{equation}
where $\sigma = 10$ is the forcing strength and $k_f = 15$ is the forcing wavenumber.
To monitor the evolution of the system, we compute integral quantities using the following convention,
\begin{equation}
    \langle f \rangle \equiv \int_\Omega f \, \mathrm{d} S = \int_0^{2 \pi} \int_0^{2 \pi} f(y, x) \, \mathrm{d}x \, \mathrm{d}y.
\end{equation}
Setting a fixed time step $\delta t = 2 \times 10^{-4}$, stationary turbulent states are obtained from random Gaussian initialization $\omega(t = 0) \sim \mathcal{N}(0, 10^{-3})$ followed by a spin-up phase until the balance of kinetic energy $E$ and enstrophy $Z$ settle to a constant value \cite{bouchet2012statistical}, with
\begin{equation}
    E = \frac{1}{2} \langle \mathbf{u}^2 \rangle = \frac{1}{2} (\langle u_x^2 \rangle + \langle u_y^2 \rangle), \quad Z = \frac{1}{2} \langle \omega^2 \rangle.
\end{equation}
Practically, we end the spin-up at $t = 1000$ where fluctuations can no longer be associated with the transient phase. Once steady-state is reached, we estimate the decorrelation time of the system by running an ensemble of perturbed simulations with $\delta \omega(t = 0) = \omega(t = 0) + 10^{-10} \omega_\text{pert}$ where $\omega_\text{pert}$ is a centered gaussian bump $e^{-10 \pi ([x - \pi]^2 + [y - \pi]^2)}$ randomly shifted in space by $(\delta x, \delta y) \in [-\pi, \pi]$. We find in Fig.~\ref{fig:qg-app} (left) that the system is fully decorrelated at $t_c \approx 25$, corresponding to $t_c \times \delta t = 125000$ time steps. The coarse-graining ratio for this experiment is fixed to $\Delta^\prime = 16$, yielding a time step $\overline{\delta t} = \Delta^\prime \delta t = 3.2 \times 10^{-3}$ for the coarse solver $g$. For illustration, we show in Fig.~\ref{fig:qg-app} examples of vorticity fields $\omega$ (center) and coarse grained vorticity fields $\bar{\omega}$ (right) at the end of the spin-up. Here, the dynamics has segregated into three stable jets of alternating directions, typical of the dynamics found in rotating objects such as the Earth's atmosphere and oceans, or in gas giants. The numerical parameters used to produce this simulation are summarized in Table~\ref{table:qg-params}.
\begin{table}
  \caption{Parameters of the numerical setup for simulation (top) and learning (bottom). 
  Note that coarse-resolution systems use the same parameters, except for grid resolution $\bar{N} = N / \Delta^\prime$, and time step $\overline{\delta t} = \Delta^{\prime} \delta t$. 
  The quantities are given in numerical (dimensionless) as directly used in the solver for reproducibility.}
  \centering
  \begin{tabular}{l l c c c c}
  \\
  \toprule
  \textbf{Name} & \textbf{Symbol} & \multicolumn{3}{c}{\textbf{Value}} \\
  \midrule
  \textit{Simulation parameters} & & \\
  \midrule
  Length of the domain & $L$ & \multicolumn{3}{c}{$2 \pi$} \\
  Linear drag & $\mu$ & \multicolumn{3}{c}{$2 \times 10^{-2}$} \\
  Kinematic viscosity & $\nu$ & \multicolumn{3}{c}{$10^{-5}$} \\
  Rossby parameter & $\beta$ & \multicolumn{3}{c}{30} \\
  Forcing amplitude & $\sigma$ & \multicolumn{3}{c}{10} \\
  Forcing wavenumber & $k_f$ & \multicolumn{3}{c}{15} \\
  Time step & $\delta t$ & \multicolumn{3}{c}{$2 \times 10^{-4}$} \\
  Number of grid points & $N$ & \multicolumn{3}{c}{2048} \\
  Grid ratio & $\Delta^\prime$ & \multicolumn{3}{c}{16} \\
  \midrule
  \textit{Learning parameters} & & $\mathcal{E}$ (sm) & $\mathcal{E}$ (lg) & $\mathcal{M}$\\
  \midrule
  Number of layers & & 8 & 8 & 5 \\
  Number of features & & 32 & 64 & 64 \\
  Kernel size & & 5 & 7 & 5 \\
  Trainable parameters & & $\sim 220$K & $\sim 1,672$K & $\sim 512$K \\
  Time steps per trajectory & $N_t$ & \multicolumn{3}{c}{$23$} \\
  Training trajectories & $N_\text{traj}$ & \multicolumn{3}{c}{$11$} \\
  \bottomrule
  \end{tabular}
  \label{table:qg-params}
\end{table}

\subsection{Learning setup}\noindent
Following previous studies \cite{frezat2022posteriori} we take trajectories of $N_t = 23$, which corresponds to $N_t \times t_c^{-1} \times \overline{\delta t} = 0.003$, i.e. 0.3\% of the decorrelation time of the system. We believe that the small-scale dynamics can be statistically representative in a single eddy turnover time,
\begin{equation}
    t_L = 2 \pi \sqrt{\frac{2}{\langle \omega^2 \rangle}},
\end{equation}
which in our configuration is estimated after spin-up to $t_L \approx 0.85$. Within this timespan, we can extract $N_\text{traj} = t_L / (N_t \times \overline{\delta t}) = 11$ continuous trajectories containing $N_t$ samples each. To generate the corresponding datasets of $N_\text{traj} \times N_t = 253$ samples, we start from the spin-up state and subsample at each iteration from $g$ for \eqref{eq:dual-emu} and every $\Delta^{\prime}$ iterations from $f$ for \eqref{eq:dual-sgs} so that these states directly correspond to one iteration performed by $g$. For the SGS model $\mathcal{M}$ (both online and emulator-based), we use a simple convolutional neural network (CNN) with $\sim 512$K learnable parameters from 5 convolution layers with kernels of size $5$ and $64$ filters each, followed by non-linear ReLu activations. This type of architecture has been shown to be able to learn extremely accurate SGS dynamics in \cite{frezat2022posteriori} for the QG system.
For the coarse dynamics represented by the neural emulator $\mathcal{E}$, we use a 8-layers residual CNN that we decline in two versions to evaluate its impact on the subsequent SGS learning step. We define a small (denoted ``sm") and a large (denoted ``lg") emulator with 32 and 64 features latent space with $5 \times 5$ and $7 \times 7$ convolution kernels for a total of $\sim 220$K and $\sim 1,672$K learnable parameters, respectively (see Table~\ref{table:qg-params} for a summary of the learning parameters and architectures used by $\mathcal{E}$ and $\mathcal{M}$).
Note that for each of these architectures, input boundaries are replicated periodically given the geometry of the domain. Using state of the art architectures for the SGS model could further improve the overall performance and might be interesting to explore. However, we may point out that the goal is not to design an optimal NN-based architecture but rather to evaluate the training strategy at the same SGS computational cost during evaluation stage. Concerning the functional loss $\ell$, we use the integral notations for both the state and subgrid losses, which are equivalent by definition to mean squared errors (MSE). For the emulator, the loss penalizes the state of the system, i.e., vorticity such that,
\begin{equation}
    \ell_\text{on}^\mathcal{E}(\{\bar{\omega}(t)\}_{t \in [t_0,t_1]}, \{\bar{\omega}_\mathcal{E}(t)\}_{t \in [t_0,t_1]}) = \frac{1}{N_t} \sum_{t \in [t_0,t_1]} \left\langle \left( \bar{\omega}_\mathcal{E}(t) - \bar{\omega}(t) \right)^2 \right\rangle,
\end{equation}
where $\bar{\omega}_\mathcal{E}(t)$ is the vorticity predicted by the emulator at time $t$. Both the small (\textit{sm}) and large (\textit{lg}) emulators are trained from 500 epochs using the AdamW optimizer with learning rate $\eta = 10^{-4}$. For the SGS models, we penalizes the kinetic energy for the state loss and the SGS term for the subgrid loss, i.e.,
\begin{align}
    \ell_\text{on}^\text{state}(\{\mathbf{u}(t)\}_{t \in [t_0,t_1]}, \{\hat{\mathbf{u}}(t)\}_{t \in [t_0,t_1]}) &= \frac{1}{N_t} \sum_{t \in [t_0,t_1]} \left\langle \left( \hat{\mathbf{u}}(t) - \bar{\mathbf{u}}(t) \right)^2 \right\rangle, \\
    \ell_\text{on}^\text{subgrid}(\{\tau_\omega(t)\}_{t \in [t_0,t_1]}, \{\hat{\omega}(t)\}_{t \in [t_0,t_1]}) &= \frac{1}{N_t} \sum_{t \in [t_0,t_1]} \left\langle \left( \mathcal{M}_\text{on}^\text{subgrid}(\hat{\omega}(t); \theta) - \tau_\omega(t) \right)^2 \right\rangle,
\end{align}
where the velocities can be easily recovered from the vorticity states through the streamfunction \eqref{eq:qg-inc}, $\mathbf{u} = (\partial_y \nabla^{-2} \omega, -\partial_x \nabla^{-2} \omega)$. The SGS models are also trained for 500 epochs using the AdamW optimizer, with learning rate $\eta = 10^{-4}$ for the state loss and $\eta = 10^{-3}$ for the subgrid loss. For additional convergence, we use a cosine decay schedule with $\alpha = 0.01$ \cite{loshchilov2017sgdr}. 
While using the MSE led to satisfying results here, we note that with the online approach, time-averaged quantities such as spectra, transfers or distributions could be used to improve performance of the neural emulator and/or SGS model further, depending on the target application.

\subsection{\textit{A priori} evaluation}
\begin{figure}[tb]
  \noindent\includegraphics[width=.9\textwidth]{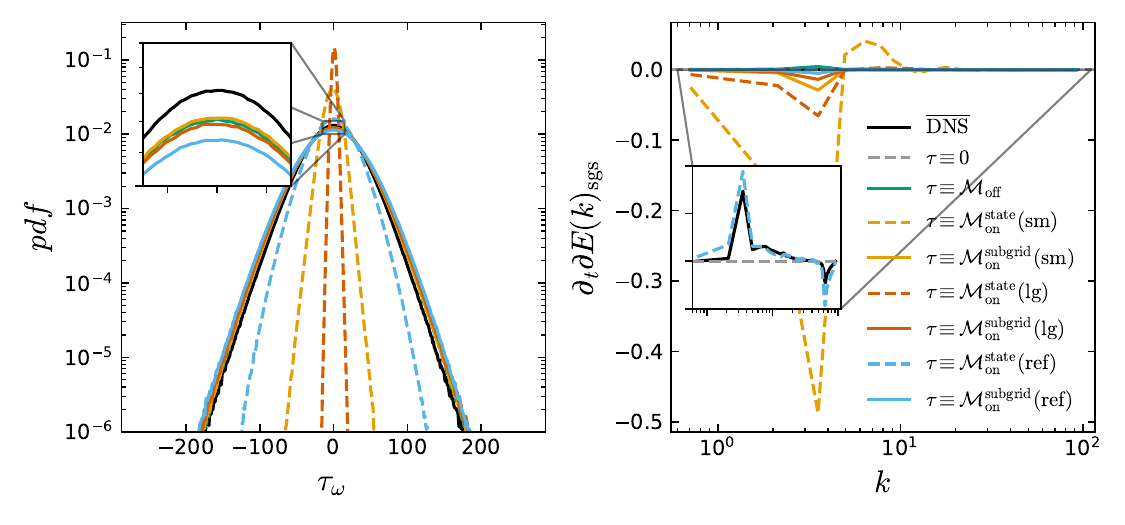}
  \caption{\textit{A priori} evaluation of the different models used in this study. The test set is composed of 2,500 filtered samples from a DNS performed after spin-up, i.e., prolonging the initial trajectory. Averaged $pdf$ of the SGS term $\tau_\omega$ (left) and energy fluxes due to the SGS term $\partial_t \partial E(k)_\text{sgs}$ (right).}
  \label{fig:qg-apriori_eval}
\end{figure}

\noindent
\label{subsec:apriori}
Once models are trained, we extend the initial simulation beyond spin-up to $t = 240$, corresponding to approximately 1.2M time steps. This evaluation window is roughly 300 times longer than the temporal horizon used in the training dataset. The resulting states $\omega(t)$ and SGS terms $\tau_\omega(t)$ form the test set for the \textit{a priori} evaluation. In addition to the reference models $\tau \equiv \mathcal{M}_\text{on} (\text{ref})$ trained using the differentiable solver $g$, and the emulator-based models $\tau \equiv \mathcal{M}_\text{on} (\text{sm,lg})$, we also include a classical ``offline" model $\tau \equiv \mathcal{M}_\text{off}$ trained with an equivalent of the subgrid loss without temporal integration during training. In Fig.~\ref{fig:qg-apriori_eval} (left), we show the probability distribution function ($pdf$) of the SGS term $\tau_\omega = \mathcal{M}(\bar{\omega})$ predicted by the different models on the test set. We can observe that the models trained with the subgrid loss (solid lines) are in close agreement with the DNS (black). To further characterize the effect of these models in \textit{a posteriori} simulations, we also examine the energy transfers induced by the SGS term,
\begin{equation}
    \frac{\partial E(k)_\text{sgs}}{\partial t} = \Re \{ \bar{\psi}^*(k) \tau_\omega(k) \}
\end{equation}
where $k$ is the (isotropic) wavenumber, $\bar{\psi}^*$ is the complex conjugate of the coarse-grained streamfunction and $\Re$ takes the real part. Note that $\tau \equiv 0$ denotes the under-resolved simulation, i.e., running on the coarse grid without any SGS model correction. 
As shown in Fig.~\ref{fig:qg-apriori_eval} (right), the models trained using the state loss $\tau \equiv \mathcal{M}_\text{on}^\text{state} (\text{sm, lg})$ incorrectly generate negative transfers (forward-scatter) at large-scale, whereas positive transfers (backscatter) is expected, as shown in the inset for the coarse-grained DNS. The models trained using the subgrid loss --including the model trained with the differentiable solver $g$-- are also generating forward-scatter at large scales, but they recover some degree of backscatter, albeit difficult to see here. Overall, the only model that accurately reproduces the subgrid energy transfers is the one trained using the solver $g$ with the state loss, $\tau \equiv \mathcal{M}_\text{on}^\text{state} (\text{ref})$.

\subsection{\textit{A posteriori} evaluation}
\label{subsec:aposteriori}
To evaluate the performance of the different models when used in a simulation, we integrate the system with each SGS model to $t = 240$, corresponding to 75,000 time steps with the coarse solver $g$.
\begin{figure}[tb]
  \noindent\includegraphics[width=.9\textwidth]{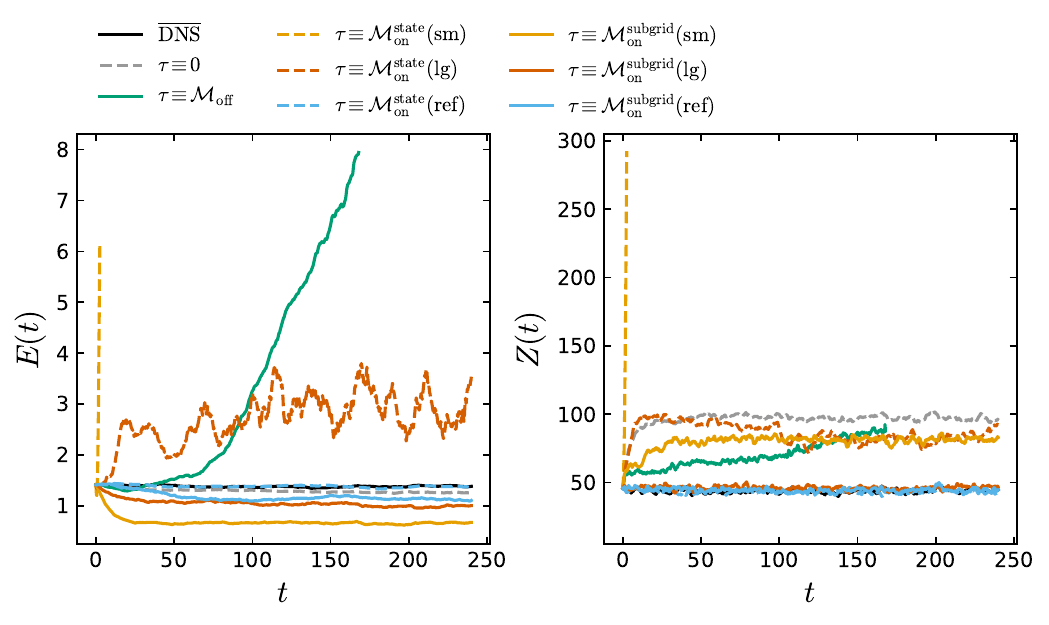}
  \caption{Evolution of kinetic energy $E(t)$ (left) and enstrophy $Z(t)$ (right) for the different models used in this study. Simulations are integrated for almost $300$ turnover times $t_L$.}
  \label{fig:qg-integrals_evolution}
\end{figure}
We first look at the temporal evolution of the two invariants of the system, i.e., kinetic energy and enstrophy in Fig.~\ref{fig:qg-integrals_evolution}. The model $\tau \equiv \mathcal{M}_\text{on}^\text{state} (\text{sm})$ exhibit a rapid growth of kinetic energy that leads to an early numerical instability. The offline-trained model $\tau \equiv \mathcal{M}_\text{off}$ also steadily accumulate energy --although at a smaller rate-- that eventually crashes the simulation at $t \approx 175$. Among the models that complete the simulation, $\tau \equiv \mathcal{M}_\text{on}^\text{state} (\text{lg})$ is notably unstable, with large oscillations from a kinetic energy level roughly 2 times larger than that of the DNS. The under-resolved simulation $\tau \equiv 0$ is relatively close to the DNS in terms of kinetic energy, but produces a large quantity of enstrophy, suggesting an accumulation of small-scale variance typically associated with the lack of spatial resolution. In contract, the enstrophy is well represented by the reference models $\tau \equiv \mathcal{M}_\text{on} (\text{ref})$ for both the state and subgrid losses, but also by $\tau \equiv \mathcal{M}_\text{on}^\text{subgrid} (\text{lg})$. Concerning the kinetic energy, the closest agreement with the DNS is obtained from the reference model trained on the state loss, while both reference and large emulator-based models trained on the subgrid loss are slightly over dissipating.
\begin{figure}[tb]
  \centering
  \noindent\includegraphics[width=.99\textwidth]{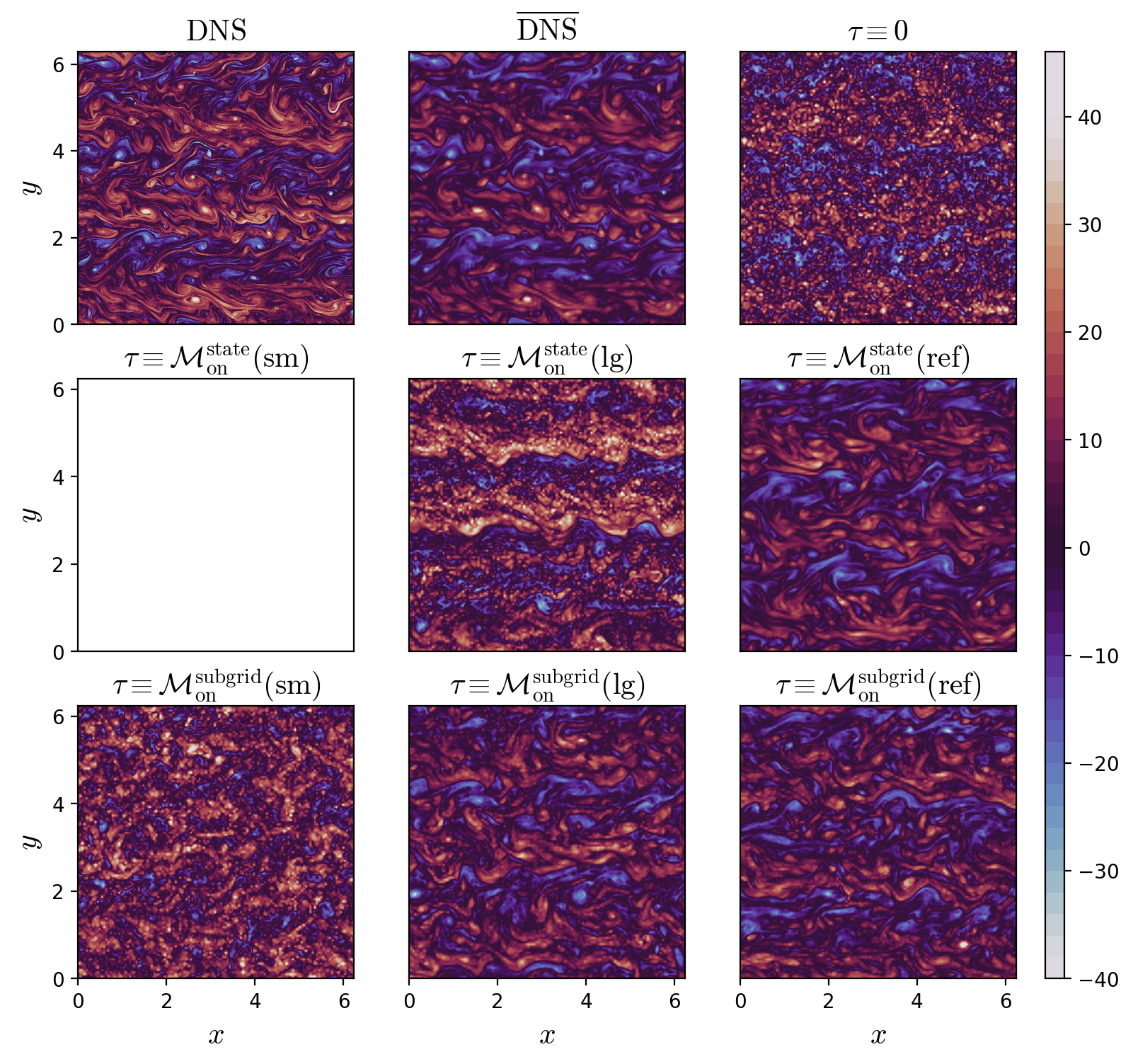}
  \caption{Vorticity fields at the end of the \textit{a posteriori} evaluation ($t = 240$) for the DNS, filtered DNS and the models in this study. Note that $\tau \equiv \mathcal{M}_\text{on}^\text{state} (\text{sm})$ was unstable and is thus shown as empty. The offline model $\tau \equiv \mathcal{M}_\text{off}$ was also omitted from this plot because it crashed during the simulation.}
  \label{fig:qg-vorticity_end}
\end{figure}
We show in Fig.~\ref{fig:qg-vorticity_end} the vorticity fields at the end of the \textit{a posteriori} integration for the DNS, filtered DNS and the different SGS models. The models with a larger enstrophy than the DNS in Fig.~\ref{fig:qg-integrals_evolution} (right), namely $\tau \equiv 0$, $\tau \equiv \mathcal{M}_\text{on}^\text{subgrid} (\text{sm})$ and $\tau \equiv \mathcal{M}_\text{on}^\text{state} (\text{lg})$ show visual signs of instabilities (fireflies). These artifacts are not observed for the reference models $\tau \equiv \mathcal{M}_\text{on} (\text{ref})$ and $\tau \equiv \mathcal{M}_\text{on}^\text{subgrid} (\text{lg})$, both of which preserve the jet structure seen in the DNS.
\begin{figure}[tb]
  \noindent\includegraphics[width=.9\textwidth]{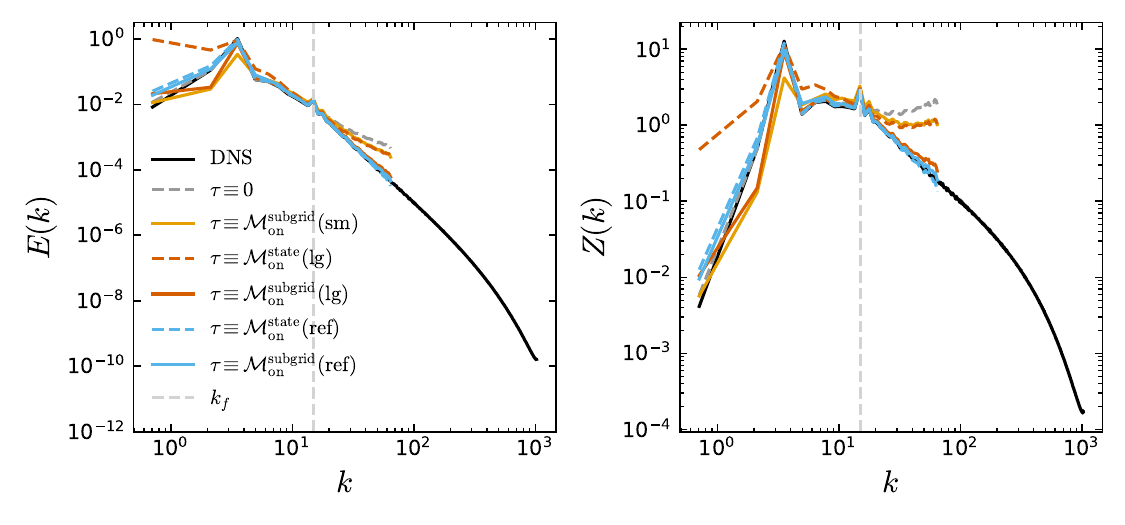}
  \caption{Kinetic energy and enstrophy spectra $E(k)$ (left) and $Z(k)$ (right) averaged over 2,500 samples on the \textit{a posteriori} evaluation horizon $t \in [0, 240]$ for the different models used in this study. The forcing wavenumber $k_f = 15$ is shown in dashed gray.}
  \label{fig:qg-spectra}
\end{figure}
From now on, we exclude the unstable models $\tau \equiv \mathcal{M}_\text{off}$ and $\tau \equiv \mathcal{M}_\text{on}^\text{state} (\text{sm})$ from the subsequent evaluation. A typical measure of the performance of a model is given by the spectra of kinetic energy and enstrophy,
\begin{equation}
    E(k) = \int_{|\mathbf{k}| = k} |\mathbf{u}(\mathbf{k})|^2 \, \mathrm{d} S(\mathbf{k}), \quad Z(k) = \int_{|\mathbf{k}| = k} |\omega(\mathbf{k})|^2 \, \mathrm{d} S(\mathbf{k}).
\end{equation}
Figure~\ref{fig:qg-spectra} shows the time-averaged spectra $E(k)$ (left) and $Z(k)$ (right). Note that the DNS spectra extends to larger wavenumbers $k$ than the different models because it resolves more spatial scales. The large scales of the kinetic energy spectrum (small $k$) are well represented by models trained with the solver $g$, but also for the under-resolved simulation $\tau \equiv 0$. However, $\tau \equiv 0$ is accumulating energy at small scale (large $k$), which is even more apparent in the enstrophy spectrum, where the scales are amplified by $k^2$, i.e., we have $Z(k) = k^2 E(k)$. While $\tau \equiv \mathcal{M}_\text{on}^\text{subgrid} (\text{sm})$ also show a similar excess, there is a relatively close agreement with the DNS for the $\tau \equiv \mathcal{M}_\text{on}^\text{subgrid} (\text{lg})$. It is important to mention that while the model trained on the state loss with the large emulator architecture does not crash compared to the one trained with the small emulator architecture, it shows a significant mismatch on the largest scales ($k = 1$), which can be attributed to the error compensation from the imperfect emulator.
\begin{figure}[tb]
  \noindent\includegraphics[width=.9\textwidth]{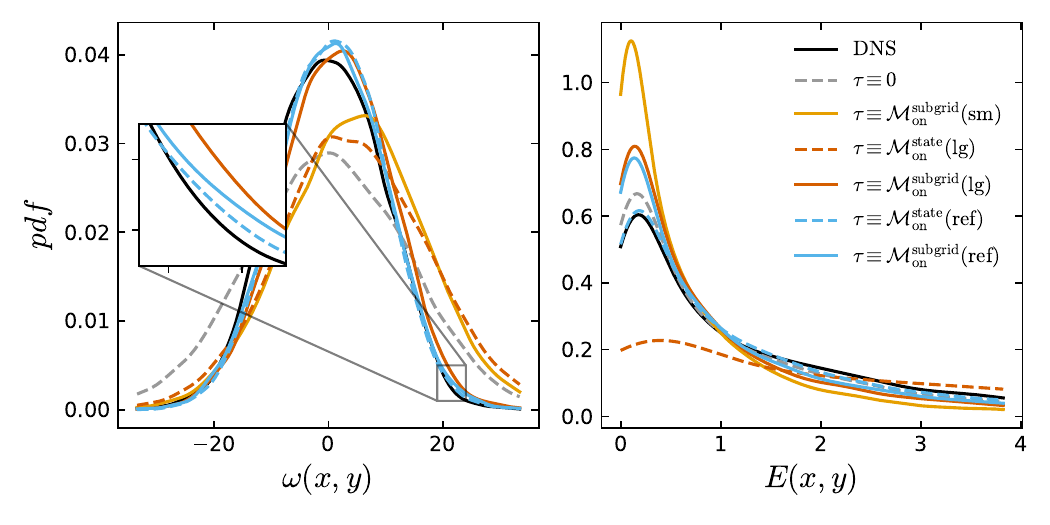}
  \caption{Probability distribution of vorticity $\omega$ and kinetic energy $E(x, y) = 1/2 (u_x^2 + u_y^2)$ at the end of the \textit{a posteriori} evaluation at $t = 240$ for the different models used in this study.}
  \label{fig:qg-distributions}
\end{figure}
These trends are consistent with the distributions of vorticity $\omega(x, y)$ and pointwise kinetic energy $E(x, y)$ shown in Fig.~\ref{fig:qg-distributions}). The under-resolved simulation $\tau \equiv 0$ has a larger vorticity spread, while $\tau \equiv \mathcal{M}_\text{on}^\text{subgrid} (\text{sm})$ and $\tau \equiv \mathcal{M}_\text{on}^\text{state} (\text{lg})$ are shifted to the right. Note that the tail of the vorticity distribution is particularly well captured by the model trained by the solver $g$ using the state loss, as highlighted in the inset. The kinetic-energy distribution is completely lost for $\tau \equiv \mathcal{M}_\text{on}^\text{state} (\text{lg})$ and has an over-estimated peak for the other models, with the exception of $\tau \equiv \mathcal{M}_\text{on}^\text{state} (\text{ref})$ that closely matches the DNS.

\subsection{Similarity metrics}
\label{subsec:similarity-metrics}
\begin{table}
  \caption{Similarity measures on the metrics used in the \textit{a posteriori} evaluation. Note that the metric is systematically calculated against the DNS with respect to the under-resolved simulation $\tau \equiv 0$.}
  \centering
  \begin{tabular}{l r r r r r r}
  \hline
   & $\mathcal{S}\{E\}$ & $\mathcal{S}\{Z\}$ & $\mathcal{S}\{E(k)\}$ & $\mathcal{S}\{Z(k)\}$ & $\mathcal{S}\{pdf(\omega)\}$ & $\mathcal{S}\{pdf(E)\}$ \\
  \hline
  $\mathcal{M}_\text{on}^\text{subgrid}(\text{sm})$ & $-6.586$ & $0.309$ & $-3.747$ & $-0.170$ & $-0.381$ & $-4.064$ \\
  $\mathcal{M}_\text{on}^\text{state}(\text{lg})$ & $-14.240$ & $0.180$ & $-8.170$ & $-0.030$ & $-0.380$ & $-15.569$ \\
  $\mathcal{M}_\text{on}^\text{subgrid}(\text{lg})$ & $-2.623$ & $0.942$ & $-1.048$ & $0.537$ & $0.592$ & $-1.808$ \\
  $\mathcal{M}_\text{on}^\text{state}(\text{ref})$ & $\mathbf{0.753}$ & $\mathbf{0.965}$ & $\mathbf{0.290}$ & $\mathbf{0.790}$ & $\mathbf{0.821}$ & $\mathbf{0.550}$ \\
  $\mathcal{M}_\text{on}^\text{subgrid}(\text{ref})$ & $-1.333$ & $\mathbf{0.966}$ & $-0.961$ & $0.543$ & $\mathbf{0.829}$ & $-1.348$ \\
  \hline
  \end{tabular}
  \label{table:qg-similarity-metrics}
\end{table}
To quantitatively compare the performance of the proposed SGS models with the under-resolved simulation $\tau \equiv 0$, we employ the similarity metrics introduced by \cite{ross2023benchmarking}. For integrated metrics in Fig.~\ref{fig:qg-integrals_evolution}, spectral metrics in Fig.~\ref{fig:qg-spectra} and distributional metrics in Fig.~\ref{fig:qg-distributions}, we define three corresponding distances,
\begin{align}
    d_t(f_{\text{sim}_1}, f_{\text{sim}_2}) &= \sqrt{\langle (f_{\text{sim}_1} - f_{\text{sim}_2})^2) \rangle}, \nonumber\\
    d_k(f_{\text{sim}_1}, f_{\text{sim}_2}) &= \sqrt{\frac{1}{|\mathbf{k}|} \sum_{k \in |\mathbf{k}|} (f_{\text{sim}_1}(k) - f_{\text{sim}_2}(k))^2}, \label{eq:distances} \\
    d_{pdf}(f_{\text{sim}_1}, f_{\text{sim}_2}) &= W_1(f_{\text{sim}_1} \sim P, f_{\text{sim}_2} \sim Q),\nonumber
\end{align}
where $f_{\text{sim}_1}$ and $f_{\text{sim}_2}$ are quantities obtained from simulations $\text{sim}_1$ and $\text{sim}_2$, respectively, and where $W_1$ refers to the Wasserstein distance defined in \eqref{eq:wasserstein}. Based on these distances, the similarity measure quantifies how much closer a given model simulation is to the filtered DNS than to the under-resolved simulation,
\begin{equation}
    \mathcal{S}\{ f_{\tau \equiv \mathcal{M}}; d\} = 1 - \frac{d(f_{\tau \equiv \mathcal{M}}, f_{\overline{\text{DNS}}})}{d(f_{\tau \equiv 0}, f_{\overline{\text{DNS}}})},
\end{equation}
where $d$ can be any of the distance defined in \eqref{eq:distances}. This similarity score is approximately 1 if the model's distance to the DNS is much smaller than that of the under-resolved simulation; it is approximately 0 if this distance is approximately equal to that of the under-resolved simulation, and is less than 0 if the distance is larger than that of the under-resolved simulation. Table~\ref{table:qg-similarity-metrics} summarizes the similarity scores across all metrics considered in the \textit{a posteriori} evaluation. The results clearly indicates that $\mathcal{M}_\text{on}^\text{state}(\text{ref})$ outperforms the other models, obtaining strictly positive similarity scores across all distance metrics. Models trained with the subgrid loss achieve positive scores for vorticity-derived metrics, such as enstrophy, but negative scores for velocity-derived metrics, including kinetic energy. A strong sensitivity to the choice of coarse-grained solver emerges for the state-loss models: the version trained with the differentiable solver $g$ performs best overall, whereas the version trained with the large neural emulator performs the worst among the stable models.

\section{Discussion}
\label{sec:discussion}
In this article, we have proposed an algorithm for training an online ML-based SGS parametrization, i.e., with \textit{a posteriori} criteria for non-differentiable numerical solvers. The core idea of our approach is to use neural emulators that represent the coarse dynamics, equivalent to the original equations in a lower dimensional space. These emulators are by definition differentiable and allow us to train the parametrization in an online setting.
The main benefit of our two-step algorithm is to avoid error compensation between the emulator and the trained ML-based parametrization using different loss targets for the SGS model and for the neural emulator, respectively. Indeed, as described in Subsection \ref{subsec:error-compensations} above, learning jointly emulator and parametrization will lead to error compensation, where the parametrization-part accommodates the imperfect nature of the emulator-part. When used in a simulation, the trained parametrization is found to add spurious large scale transfers that decreases overall performance.

Our two-step algorithm has been demonstrated on a chaotic L96 system and tested on a two-dimensional SGS modeling problem and shows performance levels close to those obtained with online learning on differentiable solvers on small-scale metrics. In a highly turbulent quasi-geostrophic setting, the model trained with our two-step algorithm is able to remain stable over temporal horizons at least 250 times larger than the training horizon. Moreover, it has a matching performance with the online model in terms of small-scale quantities, i.e., vorticity distributions and overall enstrophy spectrum.
Nonetheless, additional work will be required in order to alleviate some systematic biases in the SGS model associated with the inherently imperfect nature of the neural emulator. This error is mostly visible at large scale. We think that constraining the loss function of the neural emulator on related large scale quantities such as velocities instead of vorticity could further improve its performance, translating to a smaller bias in the learned SGS model.

This work represents a first step toward more complex neural emulators. While the experimental system used here has a considerable amount of state parameters, more work will still be required for deploying our approach into full complexity physical models. We have also seen that the SGS learning step is sensitive to the quality of the neural emulator, which could be a potential issue when emulating larger numerical systems. However, we are optimistic that improving the performance of the emulator is technically possible, as we have already seen large improvements between architectures of different complexities. Recently, some progress has been made on accurate short-range atmosphere \cite{price2025probabilistic} and ocean emulators \cite{chattopadhyay2024oceannet} using neural operators and architectures based on diffusion. While the long-term stability of these emulators is still an open question, long-horizon, but coarser-resolution emulators have also been proposed to address these issues \cite{watt2025ace2,dheeshjith2025samudra}. At this stage, it is also unclear how our algorithm compares with other existing gradient free training techniques. In particular, it would be informative to perform a comparison with the approach described in \cite{pedersen2023reliable} that proposes to use neural emulation to map model states at different future times in order to train SGS parametrizations. Systematic comparisons of the proposed approach with techniques that approximate the gradient of the loss function either by not propagating through the solver \cite{list2024temporal}, or using a first order approximation \cite{ouala2024online} would certainly benefit to the community.

Our neural emulator approach is not only relevant for SGS parametrization but also more generally for differentiable computing, with a large potential in Earth System Modeling \cite{gelbrecht2023differentiable}. Our emulation-based approach could be leveraged for similar tasks as for instance for calibrating model parameters \cite{wagner2023catke} or learning model error \cite{farchi2021using}.
Our algorithm could also provide a practical solution for deploying variational DA techniques without having to compute the adjoint of the solver \cite{carrassi2018data}. In practice, variational DA frameworks require an adjoint operator which is complicated to maintain. Adjoint-free minimization techniques have been proposed, including ensemble methods \cite{yang2015enhanced} but trades-off performance and computing time. In previous work \cite{nonnenmacher2021deep} and recent applications to different tasks, differentiable emulation opens a possible alternative approach for variational DA of numerical models.
Neural emulation therefore appears as a strategy to further explore in order to develop hybrid climate models that leverage the full potential of DP.

\appendix

\section*{Open Research Section}
The code used to generate data, train and evaluate models has been made openly available as a collection of scripts and notebooks in \url{https://github.com/hrkz/gradient-free-subgrid-neural-emulator} and archived in \citeA{hugo_frezat_2026_21493939} under the MIT licence.

\acknowledgments
This research was supported by the CNRS through the 80 PRIME project, the ANR through Melody and OceaniX. Computations were performed using GPU resources from GENCI-IDRIS. This article is a contribution to EDITO-Model Lab, a project funded by the European Union's Horizon Europe research and innovation programme under the grant No 101093293.

\bibliography{references.bib}

@article{rogallo1984numerical,
  title={Numerical simulation of turbulent flows},
  author={Rogallo, Robert S and Moin, Parviz},
  journal={Annual Review of Fluid Mechanics},
  volume={16},
  number={1},
  pages={99--137},
  year={1984},
  publisher={Annual Reviews 4139 El Camino Way, PO Box 10139, Palo Alto, CA 94303-0139, USA},
  doi={10.1146/annurev.fl.16.010184.000531}
}

@article{meneveau2000scale,
  title={Scale-invariance and turbulence models for large-eddy simulation},
  author={Meneveau, Charles and Katz, Joseph},
  journal={Annual Review of Fluid Mechanics},
  volume={32},
  number={1},
  pages={1--32},
  year={2000},
  publisher={Annual Reviews 4139 El Camino Way, PO Box 10139, Palo Alto, CA 94303-0139, USA},
  doi={10.1146/annurev.fluid.32.1.1}
}

@article{fox2019challenges,
  title={Challenges and prospects in ocean circulation models},
  author={Fox-Kemper, Baylor and Adcroft, Alistair and B{\"o}ning, Claus W and Chassignet, Eric P and Curchitser, Enrique and Danabasoglu, Gokhan and Eden, Carsten and England, Matthew H and Gerdes, R{\"u}diger and Greatbatch, Richard J and others},
  journal={Frontiers in Marine Science},
  volume={6},
  pages={65},
  year={2019},
  publisher={Frontiers Media SA},
  doi={10.3389/fmars.2019.00065},
}

@article{schneider2017earth,
  title={Earth system modeling 2.0: A blueprint for models that learn from observations and targeted high-resolution simulations},
  author={Schneider, Tapio and Lan, Shiwei and Stuart, Andrew and Teixeira, Jo{\~a}o},
  journal={Geophysical Research Letters},
  volume={44},
  number={24},
  pages={12--396},
  year={2017},
  publisher={Wiley Online Library},
  doi={10.1002/2017gl076101}
}

@article{maher2018impact,
  title={The impact of parameterized convection on climatological precipitation in atmospheric global climate models},
  author={Maher, Penelope and Vallis, Geoffrey K and Sherwood, Steven C and Webb, Mark J and Sansom, Philip G},
  journal={Geophysical Research Letters},
  volume={45},
  number={8},
  pages={3728--3736},
  year={2018},
  publisher={Wiley Online Library},
  doi={10.1002/2017gl076826}
}

@article{hewitt2020resolving,
  title={Resolving and parameterising the ocean mesoscale in {E}arth system models},
  author={Hewitt, Helene T and Roberts, Malcolm and Mathiot, Pierre and Biastoch, Arne and Blockley, Ed and Chassignet, Eric P and Fox-Kemper, Baylor and Hyder, Pat and Marshall, David P and Popova, Ekaterina and others},
  journal={Current Climate Change Reports},
  volume={6},
  pages={137--152},
  year={2020},
  publisher={Springer},
  doi={10.1007/s40641-020-00164-w}
}

@article{brunton2020machine,
  title={Machine learning for fluid mechanics},
  author={Brunton, Steven L and Noack, Bernd R and Koumoutsakos, Petros},
  journal={Annual Review of Fluid Mechanics},
  volume={52},
  pages={477--508},
  year={2020},
  publisher={Annual Reviews},
  doi={10.1146/annurev-fluid-010719-060214}
}

@article{vinuesa2022enhancing,
  title={Enhancing computational fluid dynamics with machine learning},
  author={Vinuesa, Ricardo and Brunton, Steven L},
  journal={Nature Computational Science},
  volume={2},
  number={6},
  pages={358--366},
  year={2022},
  publisher={Nature Publishing Group US New York},
  doi={10.1038/s43588-022-00264-7}
}

@article{rasp2018deep,
  title={Deep learning to represent subgrid processes in climate models},
  author={Rasp, Stephan and Pritchard, Michael S and Gentine, Pierre},
  journal={Proceedings of the National Academy of Sciences},
  volume={115},
  number={39},
  pages={9684--9689},
  year={2018},
  publisher={National Acad Sciences},
  doi={10.1073/pnas.1810286115}
}

@article{yuval2020stable,
  title={Stable machine-learning parameterization of subgrid processes for climate modeling at a range of resolutions},
  author={Yuval, Janni and O’Gorman, Paul A},
  journal={Nature communications},
  volume={11},
  number={1},
  pages={3295},
  year={2020},
  publisher={Nature Publishing Group UK London},
  doi={10.1038/s41467-020-17142-3}
}

@article{bolton2019applications,
  title={Applications of deep learning to ocean data inference and subgrid parameterization},
  author={Bolton, Thomas and Zanna, Laure},
  journal={Journal of Advances in Modeling Earth Systems},
  volume={11},
  number={1},
  pages={376--399},
  year={2019},
  publisher={Wiley Online Library},
  doi={10.1029/2018MS001472}
}

@article{guillaumin2021stochastic,
  title={Stochastic-deep learning parameterization of ocean momentum forcing},
  author={Guillaumin, Arthur P and Zanna, Laure},
  journal={Journal of Advances in Modeling Earth Systems},
  volume={13},
  number={9},
  pages={e2021MS002534},
  year={2021},
  publisher={Wiley Online Library},
  doi={10.1029/2021MS002534}
}

@article{finn2023deep,
  title={Deep learning of subgrid-scale parametrisations for short-term forecasting of sea-ice dynamics with a {M}axwell-{E}lasto-{B}rittle rheology},
  author={Finn, Tobias Sebastian and Durand, Charlotte and Farchi, Alban and Bocquet, Marc and Chen, Yumeng and Carrassi, Alberto and Dansereau, V{\'e}ronique},
  journal={EGUsphere},
  pages={1--39},
  year={2023},
  publisher={Copernicus GmbH},
  doi={10.5194/egusphere-2022-1342}
}

@article{ross2023benchmarking,
  title={Benchmarking of machine learning ocean subgrid parameterizations in an idealized model},
  author={Ross, Andrew and Li, Ziwei and Perezhogin, Pavel and Fernandez-Granda, Carlos and Zanna, Laure},
  journal={Journal of Advances in Modeling Earth Systems},
  volume={15},
  number={1},
  pages={e2022MS003258},
  year={2023},
  publisher={Wiley Online Library},
  doi={10.1029/2022MS003258}
}

@article{frezat2021physical,
  title={Physical invariance in neural networks for subgrid-scale scalar flux modeling},
  author={Frezat, Hugo and Balarac, Guillaume and Le Sommer, Julien and Fablet, Ronan and Lguensat, Redouane},
  journal={Physical Review Fluids},
  volume={6},
  number={2},
  pages={024607},
  year={2021},
  publisher={APS},
  doi={10.1103/PhysRevFluids.6.024607}
}

@article{beucler2021enforcing,
  title={Enforcing analytic constraints in neural networks emulating physical systems},
  author={Beucler, Tom and Pritchard, Michael and Rasp, Stephan and Ott, Jordan and Baldi, Pierre and Gentine, Pierre},
  journal={Physical Review Letters},
  volume={126},
  number={9},
  pages={098302},
  year={2021},
  publisher={APS},
  doi={10.1103/PhysRevLett.126.098302}
}

@article{pawar2023frame,
  title={Frame invariant neural network closures for {K}raichnan turbulence},
  author={Pawar, Suraj and San, Omer and Rasheed, Adil and Vedula, Prakash},
  journal={Physica A: Statistical Mechanics and its Applications},
  volume={609},
  pages={128327},
  year={2023},
  publisher={Elsevier},
  doi={10.1016/j.physa.2022.128327}
}

@article{seifert2020potential,
  title={Potential and limitations of machine learning for modeling warm-rain cloud microphysical processes},
  author={Seifert, Axel and Rasp, Stephan},
  journal={Journal of Advances in Modeling Earth Systems},
  volume={12},
  number={12},
  pages={e2020MS002301},
  year={2020},
  publisher={Wiley Online Library},
  doi={10.1029/2020MS002301}
}

@article{mcgibbon2019single,
  title={Single-column emulation of reanalysis of the northeast {P}acific marine boundary layer},
  author={McGibbon, Jeremy and Bretherton, Christopher},
  journal={Geophysical Research Letters},
  volume={46},
  number={16},
  pages={10053--10060},
  year={2019},
  publisher={Wiley Online Library},
  doi={10.1029/2019GL083646}
}

@article{watt2021correcting,
  title={Correcting weather and climate models by machine learning nudged historical simulations},
  author={Watt-Meyer, Oliver and Brenowitz, Noah D and Clark, Spencer K and Henn, Brian and Kwa, Anna and McGibbon, Jeremy and Perkins, W Andre and Bretherton, Christopher S},
  journal={Geophysical Research Letters},
  volume={48},
  number={15},
  pages={e2021GL092555},
  year={2021},
  publisher={Wiley Online Library},
  doi={10.1029/2021GL092555}
}

@article{frezat2022posteriori,
  title={A posteriori learning for quasi-geostrophic turbulence parametrization},
  author={Frezat, Hugo and Le Sommer, Julien and Fablet, Ronan and Balarac, Guillaume and Lguensat, Redouane},
  journal={Journal of Advances in Modeling Earth Systems},
  volume={14},
  number={11},
  pages={e2022MS003124},
  year={2022},
  publisher={Wiley Online Library},
  doi={10.1029/2022MS003124}
}

@article{sirignano2020dpm,
  title={{DPM}: A deep learning {PDE} augmentation method with application to large-eddy simulation},
  author={Sirignano, Justin and MacArt, Jonathan F and Freund, Jonathan B},
  journal={Journal of Computational Physics},
  volume={423},
  pages={109811},
  year={2020},
  publisher={Elsevier},
  doi={10.1016/j.jcp.2020.109811}
}

@article{macart2021embedded,
  title={Embedded training of neural-network subgrid-scale turbulence models},
  author={MacArt, Jonathan F and Sirignano, Justin and Freund, Jonathan B},
  journal={Physical Review Fluids},
  volume={6},
  number={5},
  pages={050502},
  year={2021},
  publisher={APS},
  doi={10.1103/PhysRevFluids.6.050502}
}

@article{duraisamy2021perspectives,
  title={Perspectives on machine learning-augmented {R}eynolds-averaged and large eddy simulation models of turbulence},
  author={Duraisamy, Karthik},
  journal={Physical Review Fluids},
  volume={6},
  number={5},
  pages={050504},
  year={2021},
  publisher={APS},
  doi={10.1103/physrevfluids.6.050504}
}

@article{sirignano2023dynamic,
  title={Dynamic deep learning {LES} closures: Online optimization with embedded {DNS}},
  author={Sirignano, Justin and MacArt, Jonathan F},
  journal={arXiv preprint},
  year={2023},
  doi={10.48550/arXiv:2303.02338}
}

@article{pahlavan2024explainable,
  title={Explainable offline-online training of neural networks for parameterizations: A {1D} gravity {W}ave-{QBO} testbed in the small-data regime},
  author={Pahlavan, Hamid A and Hassanzadeh, Pedram and Alexander, M Joan},
  journal={Geophysical Research Letters},
  volume={51},
  number={2},
  pages={e2023GL106324},
  year={2024},
  publisher={Wiley Online Library},
  doi={10.1029/2023GL106324}
}

@article{kochkov2024neural,
  title={Neural general circulation models for weather and climate},
  author={Kochkov, Dmitrii and Yuval, Janni and Langmore, Ian and Norgaard, Peter and Smith, Jamie and Mooers, Griffin and Kl{\"o}wer, Milan and Lottes, James and Rasp, Stephan and D{\"u}ben, Peter and others},
  journal={Nature},
  volume={632},
  number={8027},
  pages={1060--1066},
  year={2024},
  publisher={Nature Publishing Group UK London},
  doi={10.1038/s41586-024-07744-y}
}

@article{list2022learned,
  title={Learned turbulence modelling with differentiable fluid solvers: physics-based loss functions and optimisation horizons},
  author={List, Bj{\"o}rn and Chen, Liwei and Thuerey, Nils},
  journal={Journal of Fluid Mechanics},
  volume={949},
  pages={A25},
  year={2022},
  publisher={Cambridge University Press},
  doi={10.1017/jfm.2022.738}
}

@article{shankar2023differentiable,
  title={Differentiable physics-enabled closure modeling for {B}urgers’ turbulence},
  author={Shankar, Varun and Puri, Vedant and Balakrishnan, Ramesh and Maulik, Romit and Viswanathan, Venkatasubramanian},
  journal={Machine Learning: Science and Technology},
  volume={4},
  number={1},
  pages={015017},
  year={2023},
  publisher={IOP Publishing},
  doi={10.1088/2632-2153/acb19c}
}

@article{fan2024differentiable,
  title={Differentiable hybrid neural modeling for fluid-structure interaction},
  author={Fan, Xiantao and Wang, Jian-Xun},
  journal={Journal of Computational Physics},
  volume={496},
  pages={112584},
  year={2024},
  publisher={Elsevier},
  doi={10.1016/j.jcp.2023.112584}
}

@inproceedings{um2020solver,
  title={Solver-in-the-Loop: Learning from differentiable physics to interact with iterative {PDE}-solvers},
  author={Um, Kiwon and Brand, Robert and Fei, Yun (Raymond) and Holl, Philipp and Thuerey, Nils},
  booktitle={Advances in Neural Information Processing Systems},
  volume={33},
  pages={6111--6122},
  year={2020},
}

@inproceedings{holl2020learning,
  title={Learning to control {PDE}s with differentiable physics},
  author={Holl, Philipp and Thuerey, Nils and Koltun, Vladlen},
  booktitle={International Conference on Learning Representations},
  year={2020},
}

@inproceedings{heiden2021neuralsim,
  title={{NeuralSim}: Augmenting differentiable simulators with neural networks},
  author={Heiden, Eric and Millard, David and Coumans, Erwin and Sheng, Yizhou and Sukhatme, Gaurav S},
  booktitle={2021 IEEE International Conference on Robotics and Automation (ICRA)},
  pages={9474--9481},
  year={2021},
  organization={IEEE},
  doi={10.1109/ICRA48506.2021.9560935}
}

@inproceedings{negiar2023learning,
  title={Learning differentiable solvers for systems with hard constraints},
  author={N{\'e}giar, Geoffrey and Mahoney, Michael W and Krishnapriyan, Aditi S},
  booktitle={International Conference on Learning Representations},
  year={2023},
}

@article{kochkov2021machine,
  title={Machine learning--accelerated computational fluid dynamics},
  author={Kochkov, Dmitrii and Smith, Jamie A and Alieva, Ayya and Wang, Qing and Brenner, Michael P and Hoyer, Stephan},
  journal={Proceedings of the National Academy of Sciences},
  volume={118},
  number={21},
  pages={e2101784118},
  year={2021},
  publisher={National Acad Sciences},
  doi={10.1073/pnas.2101784118}
}

@inproceedings{takahashi2021differentiable,
  title={Differentiable fluids with solid coupling for learning and control},
  author={Takahashi, Tetsuya and Liang, Junbang and Qiao, Yi-Ling and Lin, Ming C},
  booktitle={Proceedings of the AAAI Conference on Artificial Intelligence},
  volume={35},
  number={7},
  pages={6138--6146},
  year={2021},
  doi={10.1609/aaai.v35i7.16764}
}

@article{dresdner2023learning,
  title={Learning to correct spectral methods for simulating turbulent flows},
  author={Dresdner, Gideon and Kochkov, Dmitrii and Norgaard, Peter and Zepeda-N{\'u}{\~n}ez, Leonardo and Smith, Jamie A and Brenner, Michael P and Hoyer, Stephan},
  journal={Transactions on Machine Learning Research},
  year={2023},
  url={https://openreview.net/forum?id=wNBARGxoJn}
}

@article{ramadhan2020capturing,
  title={Capturing missing physics in climate model parameterizations using neural differential equations},
  author={Ramadhan, Ali and Marshall, John and Souza, Andre and Wagner, Gregory LeClaire and Ponnapati, Manvitha and Rackauckas, Christopher},
  journal={arXiv preprint},
  year={2020},
  doi={10.48550/arXiv.2010.12559}
}

@article{gelbrecht2023differentiable,
  title={Differentiable programming for {E}arth system modeling},
  author={Gelbrecht, Maximilian and White, Alistair and Bathiany, Sebastian and Boers, Niklas},
  journal={Geoscientific Model Development},
  volume={16},
  number={11},
  pages={3123--3135},
  year={2023},
  publisher={Copernicus Publications G{\"o}ttingen, Germany},
  doi={10.5194/gmd-16-3123-2023}
}

@article{wagner2023catke,
  title={{CATKE}: a turbulent-kinetic-energy-based parameterization for ocean microturbulence with dynamic convective adjustment},
  author={Wagner, Gregory LeClaire and Hillier, Adeline and Constantinou, Navid C and Silvestri, Simone and Souza, Andre Nogueira and Burns, Keaton and Ramadhan, Ali and Hill, Christopher N and Campin, Jean-Michel and Marshall, John C and others},
  journal={Authorea Preprints},
  year={2023},
  publisher={Authorea}
}

@article{pawar2021data,
  title={Data assimilation empowered neural network parametrizations for subgrid processes in geophysical flows},
  author={Pawar, Suraj and San, Omer},
  journal={Physical Review Fluids},
  volume={6},
  number={5},
  pages={050501},
  year={2021},
  publisher={APS},
  doi={10.1103/PhysRevFluids.6.050501}
}

@article{huang2022efficient,
  title={Efficient derivative-free {B}ayesian inference for large-scale inverse problems},
  author={Huang, Daniel Zhengyu and Huang, Jiaoyang and Reich, Sebastian and Stuart, Andrew M},
  journal={Inverse Problems},
  volume={38},
  number={12},
  pages={125006},
  year={2022},
  publisher={IOP Publishing},
  doi={10.1088/1361-6420/ac99fa}
}

@article{iglesias2013ensemble,
  title={Ensemble {K}alman methods for inverse problems},
  author={Iglesias, Marco A and Law, Kody JH and Stuart, Andrew M},
  journal={Inverse Problems},
  volume={29},
  number={4},
  pages={045001},
  year={2013},
  publisher={IOP Publishing},
  doi={10.1088/0266-5611/29/4/045001}
}

@article{kovachki2019ensemble,
  title={Ensemble {K}alman inversion: a derivative-free technique for machine learning tasks},
  author={Kovachki, Nikola B and Stuart, Andrew M},
  journal={Inverse Problems},
  volume={35},
  number={9},
  pages={095005},
  year={2019},
  publisher={IOP Publishing},
  doi={10.1088/1361-6420/ab1c3a}
}

@article{lopez2022training,
  title={Training physics-based machine-learning parameterizations with gradient-free ensemble {K}alman methods},
  author={Lopez-Gomez, Ignacio and Christopoulos, Costa and Langeland Ervik, Haakon Ludvig and Dunbar, Oliver RA and Cohen, Yair and Schneider, Tapio},
  journal={Journal of Advances in Modeling Earth Systems},
  volume={14},
  number={8},
  pages={e2022MS003105},
  year={2022},
  publisher={Wiley Online Library},
  doi={10.1029/2022ms003105}
}

@article{novati2021automating,
  title={Automating turbulence modelling by multi-agent reinforcement learning},
  author={Novati, Guido and de Laroussilhe, Hugues Lascombes and Koumoutsakos, Petros},
  journal={Nature Machine Intelligence},
  volume={3},
  number={1},
  pages={87--96},
  year={2021},
  publisher={Nature Publishing Group UK London},
  doi={10.1038/s42256-020-00272-0}
}

@article{kim2022deep,
  title={Deep reinforcement learning for large-eddy simulation modeling in wall-bounded turbulence},
  author={Kim, Junhyuk and Kim, Hyojin and Kim, Jiyeon and Lee, Changhoon},
  journal={Physics of Fluids},
  volume={34},
  number={10},
  pages={},
  year={2022},
  publisher={AIP Publishing},
  doi={10.1063/5.0106940}
}

@article{bae2022scientific,
  title={Scientific multi-agent reinforcement learning for wall-models of turbulent flows},
  author={Bae, H Jane and Koumoutsakos, Petros},
  journal={Nature Communications},
  volume={13},
  number={1},
  pages={1443},
  year={2022},
  publisher={Nature Publishing Group UK London},
  doi={10.1038/s41467-022-28957-7}
}

@article{kurz2023deep,
  title={Deep reinforcement learning for turbulence modeling in large eddy simulations},
  author={Kurz, Marius and Offenh{\"a}user, Philipp and Beck, Andrea},
  journal={International Journal of Heat and Fluid Flow},
  volume={99},
  pages={109094},
  year={2023},
  publisher={Elsevier},
  doi={10.1016/j.ijheatfluidflow.2022.109094}
}

@article{wikner2020combining,
  title={Combining machine learning with knowledge-based modeling for scalable forecasting and subgrid-scale closure of large, complex, spatiotemporal systems},
  author={Wikner, Alexander and Pathak, Jaideep and Hunt, Brian and Girvan, Michelle and Arcomano, Troy and Szunyogh, Istvan and Pomerance, Andrew and Ott, Edward},
  journal={Chaos: An Interdisciplinary Journal of Nonlinear Science},
  volume={30},
  number={5},
  year={2020},
  publisher={AIP Publishing},
  doi={10.1063/5.0005541}
}

@article{arcomano2022hybrid,
  title={A hybrid approach to atmospheric modeling that combines machine learning with a physics-based numerical model},
  author={Arcomano, Troy and Szunyogh, Istvan and Wikner, Alexander and Pathak, Jaideep and Hunt, Brian R and Ott, Edward},
  journal={Journal of Advances in Modeling Earth Systems},
  volume={14},
  number={3},
  pages={e2021MS002712},
  year={2022},
  publisher={Wiley Online Library},
  doi={10.1029/2021MS002712}
}

@article{arcomano2023hybrid,
  title={A hybrid atmospheric model incorporating machine learning can capture dynamical processes not captured by its physics-based component},
  author={Arcomano, Troy and Szunyogh, Istvan and Wikner, Alexander and Hunt, Brian R and Ott, Edward},
  journal={Geophysical Research Letters},
  volume={50},
  number={8},
  pages={e2022GL102649},
  year={2023},
  publisher={Wiley Online Library},
  doi={10.1029/2022GL102649}
}

@article{cranmer2020frontier,
  title={The frontier of simulation-based inference},
  author={Cranmer, Kyle and Brehmer, Johann and Louppe, Gilles},
  journal={Proceedings of the National Academy of Sciences},
  volume={117},
  number={48},
  pages={30055--30062},
  year={2020},
  publisher={National Acad Sciences},
  doi={10.1073/pnas.1912789117}
}

@article{nonnenmacher2021deep,
  title={Deep emulators for differentiation, forecasting, and parametrization in {E}arth science simulators},
  author={Nonnenmacher, Marcel and Greenberg, David S},
  journal={Journal of Advances in Modeling Earth Systems},
  volume={13},
  number={7},
  pages={e2021MS002554},
  year={2021},
  publisher={Wiley Online Library},
  doi={10.1029/2021MS002554}
}

@article{bocquet2023surrogate,
  title={Surrogate modeling for the climate sciences dynamics with machine learning and data assimilation},
  author={Bocquet, Marc},
  journal={Frontiers in Applied Mathematics and Statistics},
  volume={9},
  pages={1133226},
  year={2023},
  publisher={Frontiers},
  doi={10.3389/fams.2023.1133226}
}

@article{baydin2018automatic,
  title={Automatic differentiation in machine learning: a survey},
  author={Baydin, Atilim Gunes and Pearlmutter, Barak A and Radul, Alexey Andreyevich and Siskind, Jeffrey Mark},
  journal={Journal of Machine Learning Research},
  volume={18},
  number={153},
  pages={1--43},
  year={2018},
}

@article{hochreiter1998vanishing,
  title={The vanishing gradient problem during learning recurrent neural nets and problem solutions},
  author={Hochreiter, Sepp},
  journal={International Journal of Uncertainty, Fuzziness and Knowledge-Based Systems},
  volume={6},
  number={02},
  pages={107--116},
  year={1998},
  publisher={World Scientific},
  doi={10.1142/s0218488598000094}
}

@book{sagaut2006large,
  title={Large Eddy Simulation for Incompressible Flows: An Introduction},
  author={Sagaut, Pierre},
  year={2005},
  publisher={Springer Science \& Business Media}
}

@article{maulik2019subgrid,
  title={Subgrid modelling for two-dimensional turbulence using neural networks},
  author={Maulik, Romit and San, Omer and Rasheed, Adil and Vedula, Prakash},
  journal={Journal of Fluid Mechanics},
  volume={858},
  pages={122--144},
  year={2019},
  publisher={Cambridge University Press},
  doi={10.1017/jfm.2018.770}
}

@article{guan2022stable,
  title={Stable a posteriori {LES} of {2D} turbulence using convolutional neural networks: Backscattering analysis and generalization to higher {R}e via transfer learning},
  author={Guan, Yifei and Chattopadhyay, Ashesh and Subel, Adam and Hassanzadeh, Pedram},
  journal={Journal of Computational Physics},
  volume={458},
  pages={111090},
  year={2022},
  publisher={Elsevier},
  doi={10.1016/j.jcp.2022.111090}
}

@article{guan2023learning,
  title={Learning physics-constrained subgrid-scale closures in the small-data regime for stable and accurate {LES}},
  author={Guan, Yifei and Subel, Adam and Chattopadhyay, Ashesh and Hassanzadeh, Pedram},
  journal={Physica D: Nonlinear Phenomena},
  volume={443},
  pages={133568},
  year={2023},
  publisher={Elsevier},
  doi={10.1016/j.physd.2022.133568}
}

@article{ouala2024online,
  title={Online calibration of deep learning sub-models for hybrid numerical modeling systems},
  author={Ouala, Said and Chapron, Bertrand and Collard, Fabrice and Gaultier, Lucile and Fablet, Ronan},
  journal={Communications Physics},
  volume={7},
  number={1},
  pages={402},
  year={2024},
  publisher={Nature Publishing Group UK London},
  doi={10.1038/s42005-024-01880-7}
}

@inproceedings{lorenz1996predictability,
  title={Predictability: A problem partly solved},
  author={Lorenz, Edward N},
  booktitle={Proc. Seminar on predictability},
  volume={1},
  number={1},
  pages={1--18},
  year={1996},
  organization={Reading},
  doi={10.1017/CBO9780511617652.004}
}

@article{balwada2024learning,
  title={Learning Machine Learning with {L}orenz-96},
  author={Balwada, Dhruv and Abernathey, Ryan and Acharya, Shantanu and Adcroft, Alistair and Brener, Judith and Balaji, V and Bhouri, Mohamed Aziz and Bruna, Joan and Bushuk, Mitch and Chapman, Will and others},
  journal={Journal of Open Source Education},
  volume={7},
  number={82},
  pages={241},
  year={2024},
  doi={10.21105/jose.00241}
}

@article{lesieur1996new,
  title={New trends in large-eddy simulations of turbulence},
  author={Lesieur, Marcel and Metais, Olivier},
  journal={Annual Review of Fluid Mechanics},
  volume={28},
  number={1},
  pages={45--82},
  year={1996},
  publisher={Annual Reviews 4139 El Camino Way, PO Box 10139, Palo Alto, CA 94303-0139, USA},
  doi={10.1146/annurev.fl.28.010196.000401}
}

@article{piomelli1991subgrid,
  title={Subgrid-scale backscatter in turbulent and transitional flows},
  author={Piomelli, Ugo and Cabot, William H and Moin, Parviz and Lee, Sangsan},
  journal={Physics of Fluids A: Fluid Dynamics},
  volume={3},
  number={7},
  pages={1766--1771},
  year={1991},
  publisher={American Institute of Physics},
  doi={10.1063/1.857956}
}

@article{schumann1995stochastic,
  title={Stochastic backscatter of turbulence energy and scalar variance by random subgrid-scale fluxes},
  author={Schumann, Ulrich},
  journal={Proceedings of the Royal Society of London. Series A: Mathematical and Physical Sciences},
  volume={451},
  number={1941},
  pages={293--318},
  year={1995},
  publisher={The Royal Society London},
  doi={10.1098/rspa.1995.0126}
}

@article{liu2011modification,
  title={Modification of {S}palart--{A}llmaras model with consideration of turbulence energy backscatter using velocity helicity},
  author={Liu, Yangwei and Lu, Lipeng and Fang, Le and Gao, Feng},
  journal={Physics Letters A},
  volume={375},
  number={24},
  pages={2377--2381},
  year={2011},
  publisher={Elsevier},
  doi={10.1016/j.physleta.2011.05.023}
}

@article{danilov2019toward,
  title={Toward consistent subgrid momentum closures in ocean models},
  author={Danilov, Sergey and Juricke, Stephan and Kutsenko, Anton and Oliver, Marcel},
  journal={Energy Transfers in Atmosphere and Ocean},
  pages={145--192},
  year={2019},
  publisher={Springer},
  doi={10.1007/978-3-030-05704-6_5}
}

@article{fox2008can,
  title={Can large eddy simulation techniques improve mesoscale rich ocean models?},
  author={Fox-Kemper, B and Menemenlis, D},
  journal={Washington DC American Geophysical Union Geophysical Monograph Series},
  volume={177},
  pages={319--337},
  year={2008},
  doi={10.1029/177GM19}
}

@article{jansen2015energy,
  title={Energy budget-based backscatter in an eddy permitting primitive equation model},
  author={Jansen, Malte F and Held, Isaac M and Adcroft, Alistair and Hallberg, Robert},
  journal={Ocean Modelling},
  volume={94},
  pages={15--26},
  year={2015},
  publisher={Elsevier},
  doi={10.1016/j.ocemod.2015.07.015}
}

@article{juricke2019ocean,
  title={Ocean kinetic energy backscatter parametrizations on unstructured grids: Impact on mesoscale turbulence in a channel},
  author={Juricke, Stephan and Danilov, Sergey and Kutsenko, Anton and Oliver, Marcel},
  journal={Ocean Modelling},
  volume={138},
  pages={51--67},
  year={2019},
  publisher={Elsevier},
  doi={10.1016/j.ocemod.2019.03.009}
}

@article{juricke2020ocean,
  title={Ocean kinetic energy backscatter parametrization on unstructured grids: Impact on global eddy-permitting simulations},
  author={Juricke, Stephan and Danilov, Sergey and Koldunov, Nikolay and Oliver, Marcel and Sidorenko, Dmitry},
  journal={Journal of Advances in Modeling Earth Systems},
  volume={12},
  number={1},
  pages={e2019MS001855},
  year={2020},
  publisher={Wiley Online Library},
  doi={10.1029/2019MS001855}
}

@article{frederiksen2012stochastic,
  title={Stochastic subgrid parameterizations for atmospheric and oceanic flows},
  author={Frederiksen, Jorgen S and O'Kane, Terence J and Zidikheri, Meelis J},
  journal={Physica Scripta},
  volume={85},
  number={6},
  pages={068202},
  year={2012},
  publisher={IOP Publishing},
  doi={10.1088/0031-8949/85/06/068202}
}

@incollection{leonard1975energy,
  title={Energy cascade in large-eddy simulations of turbulent fluid flows},
  author={Leonard, Athony},
  booktitle={Advances in Geophysics},
  volume={18},
  pages={237--248},
  year={1975},
  publisher={Elsevier},
  doi={10.1016/S0065-2687(08)60464-1}
}

@book{canuto2007spectral,
  title={Spectral methods: Fundamentals in single domains},
  author={Canuto, Claudio and Hussaini, M Yousuff and Quarteroni, Alfio and Zang, Thomas A},
  year={2007},
  publisher={Springer Science \& Business Media}
}

@article{boscarino2013implicit,
  title={Implicit-explicit {R}unge--{K}utta schemes for hyperbolic systems and kinetic equations in the diffusion limit},
  author={Boscarino, Sebastiano and Pareschi, Lorenzo and Russo, Giovanni},
  journal={SIAM Journal on Scientific Computing},
  volume={35},
  number={1},
  pages={A22--A51},
  year={2013},
  publisher={SIAM},
  doi={10.1137/110842855}
}

@article{guan2024online,
  title={Online learning of eddy-viscosity and backscattering closures for geophysical turbulence using ensemble {K}alman inversion},
  author={Guan, Yifei and Hassanzadeh, Pedram and Schneider, Tapio and Dunbar, Oliver and Huang, Daniel Zhengyu and Wu, Jinlong and Lopez-Gomez, Ignacio},
  journal={arXiv preprint},
  year={2024},
  doi={10.48550/arXiv.2409.04985}
}

@article{bouchet2012statistical,
  title={Statistical mechanics of two-dimensional and geophysical flows},
  author={Bouchet, Freddy and Venaille, Antoine},
  journal={Physics Reports},
  volume={515},
  number={5},
  pages={227--295},
  year={2012},
  publisher={Elsevier},
  doi={10.1016/j.physrep.2012.02.001}
}

@inproceedings{loshchilov2017sgdr,
  title={{SGDR}: Stochastic gradient descent with warm restarts},
  author={Loshchilov, Ilya and Hutter, Frank},
  booktitle={International Conference on Learning Representations},
  year={2017},
  url={https://openreview.net/forum?id=Skq89Scxx}
}

@article{price2025probabilistic,
  title={Probabilistic weather forecasting with machine learning},
  author={Price, Ilan and Sanchez-Gonzalez, Alvaro and Alet, Ferran and Andersson, Tom R and El-Kadi, Andrew and Masters, Dominic and Ewalds, Timo and Stott, Jacklynn and Mohamed, Shakir and Battaglia, Peter and others},
  journal={Nature},
  volume={637},
  number={8044},
  pages={84--90},
  year={2025},
  publisher={Nature Publishing Group UK London},
  doi={10.1038/s41586-024-08252-9}
}

@article{chattopadhyay2024oceannet,
  title={{OceanNet}: a principled neural operator-based digital twin for regional oceans},
  author={Chattopadhyay, Ashesh and Gray, Michael and Wu, Tianning and Lowe, Anna B and He, Ruoying},
  journal={Scientific Reports},
  volume={14},
  number={1},
  pages={21181},
  year={2024},
  publisher={Nature Publishing Group UK London},
  doi={10.1038/s41598-024-72145-0}
}

@article{watt2025ace2,
  title={{ACE2}: accurately learning subseasonal to decadal atmospheric variability and forced responses},
  author={Watt-Meyer, Oliver and Henn, Brian and McGibbon, Jeremy and Clark, Spencer K and Kwa, Anna and Perkins, W Andre and Wu, Elynn and Harris, Lucas and Bretherton, Christopher S},
  journal={npj Climate and Atmospheric Science},
  volume={8},
  number={1},
  pages={205},
  year={2025},
  publisher={Nature Publishing Group UK London},
  doi={10.1038/s41612-025-01090-0}
}

@article{dheeshjith2025samudra,
  title={{Samudra}: An {AI} global ocean emulator for climate},
  author={Dheeshjith, Surya and Subel, Adam and Adcroft, Alistair and Busecke, Julius and Fernandez-Granda, Carlos and Gupta, Shubham and Zanna, Laure},
  journal={Geophysical Research Letters},
  volume={52},
  number={10},
  pages={e2024GL114318},
  year={2025},
  publisher={Wiley Online Library},
  doi={10.1029/2024GL114318}
}

@inproceedings{pedersen2023reliable,
  title={Reliable coarse-grained turbulent simulations through combined offline learning and neural emulation},
  author={Pedersen, Christian and Zanna, Laure and Bruna, Joan and Perezhogin, Pavel},
  booktitle={ICML2023 Workshop on the Synergy of Scientific and Machine Learning Modeling},
  year={2023},
}

@article{list2024temporal,
  title={How temporal unrolling supports neural physics simulators},
  author={List, Bj{\"o}rn and Chen, Liwei and Bali, Kartik and Thuerey, Nils},
  journal={arXiv preprint},
  year={2024},
  doi={10.48550/arXiv.2402.12971}
}

@article{farchi2021using,
  title={Using machine learning to correct model error in data assimilation and forecast applications},
  author={Farchi, Alban and Laloyaux, Patrick and Bonavita, Massimo and Bocquet, Marc},
  journal={Quarterly Journal of the Royal Meteorological Society},
  volume={147},
  number={739},
  pages={3067--3084},
  year={2021},
  publisher={Wiley Online Library},
  doi={10.1002/qj.4116}
}

@article{carrassi2018data,
  title={Data assimilation in the geosciences: An overview of methods, issues, and perspectives},
  author={Carrassi, Alberto and Bocquet, Marc and Bertino, Laurent and Evensen, Geir},
  journal={Wiley Interdisciplinary Reviews: Climate Change},
  volume={9},
  number={5},
  pages={e535},
  year={2018},
  publisher={Wiley Online Library},
  doi={10.1002/wcc.535}
}

@article{yang2015enhanced,
  title={Enhanced ensemble-based {4DVar} scheme for data assimilation},
  author={Yang, Yin and Robinson, Cordelia and Heitz, Dominique and M{\'e}min, Etienne},
  journal={Computers \& Fluids},
  volume={115},
  pages={201--210},
  year={2015},
  publisher={Elsevier},
  doi={10.1016/j.compfluid.2015.03.025}
}

@software{hugo_frezat_2026_21493939,
  author={Hugo Frezat},
  title={hrkz/gradient-free-subgrid-neural-emulator},
  month=aug,
  year=2026,
  publisher={Zenodo},
  doi={10.5281/zenodo.21493939},
  url={https://doi.org/10.5281/zenodo.21493939},
}

\end{document}